\documentclass[
preprint, %% Preprint: PrimaryVer, onecolumn, 12pt, increases line spacing %%%%
secnumarabic,amssymb, nobibnotes, aps, prd
,superscriptaddress,
nofootinbib,
]{revtex4-1}

\usepackage[ddmmyyyy]{datetime} %Displaying last compiled time 
\usepackage{amsmath} % Package for \eqref
\usepackage{xcolor} % Package for \textcolor
\usepackage{cancel} % Package for \cancelto{0}{x}
\usepackage{graphicx}
\usepackage{comment} %SA added 14/01/2019
\usepackage{soul,ulem} %For striking out - PK: March 28, 2019 
\usepackage{multirow,slashbox}
\soulregister\cite7
\soulregister\ref7 
\soulregister\eqref7 %To abose soul to make \hl through cite, ref, and eqref % https://tex.stackexchange.com/questions/139463/how-to-make-hl-highlighting-to-automatically-place-incompatible-commands-in
\usepackage{bbding}
\usepackage{pifont}
\usepackage{wasysym}
\usepackage{marginnote} %\marginpar{xyz} \marginnote{text}[3cm] \reversemarginpar
\usepackage[colorinlistoftodos
, textsize=tiny]{todonotes} %\todo{text}
\usepackage{textcomp} % added by PK on Oct 9 2019 for the  \textrightarrow \textdownarrow 
\usepackage{lineno} % added by PK on Oct 18 for the line number
\definecolor{brightpink}{rgb}{1.0, 0.0, 0.5}%brightpink
\definecolor{mygreen1}{RGB}{25,116,5} %green
\definecolor{mygreen2}{RGB}{135,10,190} % purple {135,10,110}

\begin{document}

%\title{Observational predictions of \\
%the viable generalized scalar-tensor theory}%
\title{Cosmological evolution of viable models in \\
the generalized scalar-tensor theory} 
\author{Shun Arai}
\email{arai.shun@a.mbox.nagoya-u.ac.jp}
\affiliation {Department of Physics and Astrophysics, Nagoya University, Nagoya 464-8602, Japan}
\affiliation {Institute for Astronomy, University of Edinburgh,
Royal Observatory, Blackford Hill, Edinburgh, EH9 3HJ, U.K.}

 \author{Purnendu Karmakar}
  \email{purnendu.karmakar@pd.infn.it}
  \affiliation{Physics Department and INFN, Universit\`a di Roma “La Sapienza”, Ple Aldo Moro 2, 00185, Rome, Italy}
  
\author {Atsushi Nishizawa}
\email{anishi@resceu.s.u-tokyo.ac.jp}
\affiliation {Research Center for the Early Universe (RESCEU), School of Science, The University of Tokyo, Tokyo 113-0033, Japan}
\affiliation {Kobayashi-Maskawa Institute for the Origin of Particles and the Universe, Nagoya University, Nagoya 464-8602, Japan}

%%%%%%%%% Abstract %%%%%%%%%%
\begin{abstract}
We investigate the parameter distributions of the viable generalized scalar-tensor theory with conventional dust matter after GW170817 in a model-independent way. We numerically construct the models by computing the time evolution of a scalar field, which leads to a positive definite second-order Hamiltonian and are consistent with the observed Hubble parameter. We show the model parameter distributions in the degenerate higher-order scalar-tensor (DHOST) theory, and its popular subclasses, e.g., Horndeski and GLPV theories, etc. We find that 1) the Planck mass run rate, $\alpha_M$, is insensitive to distinguish the theories, 2) the kinetic-braiding parameter, $\alpha_B$, marginally discriminates the models from those of the Horndeski theory in some range, 3) the parameters for the higher-order theories, $\alpha_H$ and $\beta_1$, are relatively smaller in magnitude (by several factors) than $\alpha_M$ and $\alpha_B$, but can still be used for discriminating the theories except for the GLPV theory. Based on the above three facts, we propose a minimal set of parameters that sensibly distinguishes the subclasses of DHOST theories, ($\alpha_M$, $\alpha_B-\alpha_M/2$, $\beta_1$).
\end{abstract}

\date{\today} % Dated: November 16, 2018 at 01:27 of Japan
\maketitle
% \tableofcontents % Table of the contents

%%%%%%%%%%%%%%%%%%%%%%%%%%%%%%%%%%%%%%%%
%%%%% Main Content starts from here %%%%

\section{Introduction}\label{sec:introduction}

As observed, our Universe is currently undergoing the phase of the late-time accelerated expansion \cite{Riess:1998cb, Perlmutter:1998np}. The challenge is finding the appropriate model or theory for explaining all observed phenomenons concurrently with theoretical consequences. The general relativity (GR) with the addition of cosmological constant, $\Lambda$, and cold dark matter can successfully explain the majority of the cosmological observational data with a minimum set of six parameters \cite{Adam:2015rua, Ade:2015xua}, called Lambda Cold Dark Matter ($\Lambda$CDM) model. The $\Lambda$CDM model accommodates the cosmological observations with high precision that we call it the standard model of cosmology. 

In the $\Lambda$CDM model, the tiny cosmological constant is responsible for explaining the present cosmic acceleration, which introduces the well-known cosmological constant (CC) problems (see review \cite{Weinberg:1988cp, Bull:2015stt}). The possible alternative explanations to the cosmic acceleration are i) replacing the cosmological constant by a dynamical scalar field as dark energy (DE) (e.g. quintessence \cite{Caldwell:1997ii}, k-essence \cite{Chiba:1999ka,ArmendarizPicon:2000ah}), or ii) introducing a modified gravitational coupling which differs from GR at cosmological distance, known as the modified gravity (MG) \cite{Tsujikawa:2010zza,Nojiri:2010wj,Clifton:2011jh,Joyce:2016vqv,Nojiri:2017ncd}. 

The degenerate higher-order scalar-tensor (DHOST) theory is claimed to be the most general class of a scalar-tensor theory with a propagating scalar and two tensor degrees of freedom given under the general-covariance \cite{Langlois:2015cwa,Crisostomi:2016czh,Achour:2016rkg,Motohashi:2016ftl,BenAchour:2016fzp} \cite{Kobayashi:2019hrl} for the review. Many modified gravity models, including the Brans-Dicke theory \cite{Brans:1961sx}, $f(R)$ gravity \cite{DeFelice:2010aj,Nojiri:2010wj}, covariant Galileon \cite{Deffayet:2009wt,Neveu:2013mfa,Neveu:2016gxp}, Horndeski \cite{Horndeski:1974wa,Kobayashi:2011nu}, transforming gravity \cite{Zumalacarregui:2013pma}, and GLPV theory \cite{Gleyzes:2014dya}, are subsets of the DHOST theory. Therefore it can be used as a generalized framework for testing gravity.

Since GR is well tested at the small scales, the scalar interactions on small scales should be suppressed for the generalized scalar-tensor theories, called the screening mechanism. It is known that in many of the viable DHOST theories, the Vainshtein screening breaks down inside the matter sources, i.e., the gravitational laws are modified \cite{Crisostomi:2017lbg,Langlois:2017dyl,Bartolo:2017ibw,Dima:2017pwp,Hirano:2019scf,Crisostomi:2019yfo}. This is a distinguished feature of the DHOST theory \cite{Crisostomi:2018bsp,Frusciante:2018tvu,Hirano:2019nkz} that is not seen in the Horndeski theory \cite{Kimura:2011dc,Narikawa:2013pjr,Koyama:2013paa,Kase:2013uja} even at small scales. %\PK{If we explain $c_g, c$ here, we would also have to talk on GW, which is done in the next paragraph. Possibility 1) Perhaps we may erase the highlighted part, and add `viable'. Possibility 2) Erase the entire paragraph and add two lines in the end of next GW paragraph "It is known that ... even at small scales."}

On the other hand, modifications of gravity often either change the speed or amplitude damping of gravitational waves (GW) propagation, or both \cite{Nishizawa:2014zna}. Therefore a GW is a new powerful tool probing the modified gravity models. Lately, LIGO and VIRGO have detected a lot of binary black holes merging events at a cosmological distance and several binary neutron star mergers, including GW170817 \cite{TheLIGOScientific:2017qsa}. Fermi and the International Gamma-Ray Astrophysics Laboratory have detected the associated electromagnetic transient, the gamma-ray burst GRB170817A \cite{Monitor:2017mdv}. As predicted before \cite{Nishizawa:2014zna,Lombriser:2015sxa}, these events, i.e., GW170817 and GRB170817A, together allow us to put the constraint on the speed of GW propagation, $c_g$, with respect to the speed of light, $c$, $|c_g^2/c^2 - 1| \lesssim 10^{-15}$ \cite{TheLIGOScientific:2017qsa, Monitor:2017mdv}. A large class of modified gravity models subclassed in Horndeski, GLPV, and DHOST theories which changes the speed of the GW propagation has been tightly constrained from the BNS merging observation \cite{Ezquiaga:2017ekz, Sakstein:2017xjx,Creminelli:2017sry,Baker:2017hug, Jain:2015edg, Arai:2017hxj,Gumrukcuoglu:2017ijh,Oost:2018tcv,Gong:2018cgj,Gong:2018vbo}, and left the Horndeski and GLPV theories with three and the DHOST theory with four arbitrary functions. 

In the effective field theory (EFT) description of the DHOST theory \cite{Langlois:2017mxy}, that gives the deviation from the $\Lambda$CDM model at linear level, the DHOST theory is expressed in terms of six time-dependent parameters in the linear perturbations, i.e., $\alpha_{M,B,K,H,T}$ and $\beta_1$
\footnote{Another EFT parameter, $\alpha_L$, represents the detuning of the extrinsic curvature term. The condition for gradient instability-free DHOST theories is $\alpha_L=0$ \cite{Langlois:2017mxy}.}.
The condition $c_g=c$ constrains the tensor speed alteration parameter, $\alpha_T$ tightly. The rest five parameters, $\alpha_{M,B,K,H}$ and $\beta_1$, are the measures to the deviation from the $\Lambda$CDM model. The constraint on the EFT parameters from the decay of gravitational waves into dark energy fluctuations has been demonstrated in \cite{Creminelli:2018xsv,Creminelli:2019nok}. However, a model-independent investigation of DHOST theory consistent with the expansion history of the universe has not been studied yet.

The knowledge on the distributions and correlations of the free functions of the DHOST theory or its EFT parameters provides intimate knowledge, for analysing or forecasting against the observation, especially cosmological surveys on the large scale structure \cite{Hildebrandt:2016iqg,Troxel:2017xyo,Hikage:2018qbn, Amendola:2012ys,Abate:2012za,Spergel:2013tha} and gravitational waves \cite{AmaroSeoane:2012km,Sathyaprakash:2012jk,Sato:2017dkf,Reitze:2019iox}. In contrast to that, the model distributions in the parameter space of the Horndeski theory have been studied in \cite{Gleyzes:2017kpi,Arai:2017hxj,Noller:2018eht, Nishizawa:2019rra}, the whole parameter space of DHOST theory has not been yet investigated.

Therefore, in this paper, we will mainly focus on showing the correlations of the linear EFT parameters of the DHOST theory ($c_g=c$) in a model-independent way. We briefly summarize the DHOST framework after GW170817 in Sec.~\ref{sec:action}, and introduce the EFT parametrization of the DHOST theory in Sec.~\ref{sec:prams}. We explain the methodology and the approximations used in our analysis in Sec.~\ref{sec:setup}. The distributions of the models in the space of the EFT parameters and the distinguishability of the subdivided theories in DHOST theory are presented in Sec.~\ref{sec:modeldist}.

We use the metric signature $(-,+,+,+)$, and set the speed of light to unity, $c=1$. Greek indices run from 0 to 3.

%%%%%%%%%%%%%%%%%%%%%%%%%%%%%%%%%%%%%%
\section{DHOST theory after GW170817} \label{sec:action}

Let us consider the general DHOST action containing a metric tensor ($g_{\mu\nu}$) and a single scalar field ($\phi$) \cite{Langlois:2015cwa,BenAchour:2016fzp,Crisostomi:2016czh}, 
\begin{eqnarray}\label{ac:host}
 S&=& \int d^4x\,  \sqrt{-g}\,  \mathcal{L}\,, \label{ac:total}
\end{eqnarray}

where the DHOST Lagrangian, $\mathcal{L}$ is defined as the sum of the following four parts,
\begin{equation}\label{l:total}
\mathcal{L} = \mathcal{L}_{\rm g} + \mathcal{L}_\phi + \mathcal{L}_{\rm oth} + \mathcal{L}_{\rm m} \,,
\end{equation}
with
\begin{eqnarray} \label{l:exp}
\mathcal{L}_{\rm g} &\equiv& F(\phi,X) R\,,\\
\mathcal{L}_{\phi}&\equiv&\sum_{i=1}^5 A_{i} (\phi,X)  \mathcal{L}_{i}\,,
\\
\mathcal{L}_{\rm oth}&\equiv&P(\phi,X) + Q (\phi,X) \Box \phi \,,
\end{eqnarray}
where $X \equiv \nabla_\mu \phi \nabla^\mu \phi$, and $F, P, Q, A_i$ are the arbitrary functions of $\phi$ and $X$. The $\mathcal{L}_\phi$ contains all possible contractions of a scalar field of the quadratic polynomial degree in second-order derivatives of the scalar field, i.e., $\phi_{\mu\nu}$ with
\begin{equation}
\begin{split}
& \mathcal{L}_1 = \phi_{\mu \nu} \phi^{\mu \nu} \,, \quad
\mathcal{L}_2 =(\Box \phi)^2 \,, \quad
\mathcal{L}_3 = (\Box \phi) \phi^{\mu} \phi_{\mu \nu} \phi^{\nu} \,,  \\
& \mathcal{L}_4 =\phi^{\mu} \phi_{\mu \rho} \phi^{\rho \nu} \phi_{\nu} \,, \quad
\mathcal{L}_5= (\phi^{\mu} \phi_{\mu \nu} \phi^{\nu})^2\,.
\end{split}
\end{equation}

The matter Lagrangian, $\mathcal{L}_{\rm m}$, is assumed to be minimally coupled to the metric, $g_{\mu\nu}$. Here, we are using the short hand notations $\phi_{\mu} = \nabla_{\mu} \phi$, $\phi_{\mu\nu} = \nabla_{\nu}\nabla_{\mu} \phi$. 
The Lagrangian $L_{\rm oth}$ also known as kinetic gravity braiding \cite{Deffayet:2010qz, Kobayashi:2010cm}. Note that, one can recover the standard GR by setting $F=1/16 \pi G$, and $P=Q=A_i = 0$.

Among the degenerate classes of DHOST theory, only the dubbed class Ia does not suffer from the gradient instability \cite{Langlois:2017mxy}. One could enlarge DHOST theory by adding the cubic, quartic or quintic dependencies on $\phi_{\mu\nu}$ to the action~\eqref{l:total}. For fulfilling the constraint on the speed of the GW, $c_g^2 = 1$, independent of any background, $A_1=0$ for the quadratic polynomial degree of DHOST theory given in Eq.~\eqref{l:total}, and all cubic or higher polynomial degrees should be vanished, hence not discussed here. Since we are interested in the theories $c_g^2=1$ in this paper, hereafter, we refer ${\rm DHOST}_{c_g^2=1}$, ${\rm GLPV}_{c_g^2=1}$, ${\rm Horndeski}_{c_g^2=1}$ just as DHOST, GLPV, and Horndeski theory respectively. The degeneracy conditions, which ensures the absence of the Ostrogradsky ghost of the class Ia DHOST theory after the GW170817 event are
\begin{equation}\label{A_cg=1}
\begin{split}
& A_1 = -A_2=0\,,  \\
& A_4=\frac{1}{8F}\left[ 48 {F_X}^2 -8(F-X{F_X}) A_3-X^2 A_3^2\right] \,, \\
& A_5 =\frac{1}{2 F}\left(4{F_X}+X A_3\right) A_3\,,
\end{split}
\end{equation}
where ${F_X} = {\partial F}/{\partial X}$.

The corresponding Lagrangian of the Class Ia DHOST theory after GW170817 event is 
\begin{equation}
\label{ac:dhost:gw}
\begin{split}
& L^{_{\rm DHOST}}_{c_g^2=1}=  P + Q\,  \Box\phi +  F \,  R +  A_3\phi^\mu \phi^\nu \phi_{\mu \nu} \Box \phi  \\
&+\frac{1}{8F} \bigg(48 {F_X}^2 -8(F-X{F_X}) A_3-X^2 A_3^2 \bigg) \phi^\mu \phi_{\mu \nu} \phi_\lambda \phi^{\lambda \nu} \\
&+\frac{1}{2 F}\left(4{F_X}+X A_3\right) A_3(\phi_\mu \phi^{\mu \nu } \phi_\nu)^2 \;. 
\end{split}
\end{equation}
GLPV theory: Viable $(c_g^2 = 1)$ GLPV theory can be identified within the above viable DHOST theory with the following mapping
\begin{eqnarray}
A_3 = - A_4\,, \quad 4{F_X}+X A_3 =0\,, \quad A_5 = 0\,, \label{e:GLPVlimit}
\end{eqnarray}
which can be visualize in GLPV notation \cite{Gleyzes:2014dya} as 
\begin{eqnarray*}%\label{e:map:GLPV:GW170817}
F = G_4\,,\quad  A_3 = - A_4 =-\frac{4}{X}G_{4X} \,, \quad A_5 = 0\,. 
\end{eqnarray*}

We call Eq. \eqref{e:GLPVlimit} as GLPV limit in rest of the article. 

Horndesky theory: Viable $(c_g^2 = 1)$ Horndeski theory can be seen as a subclass of the viable GLPV theory from Eq.~\eqref{e:GLPVlimit} by further setting
\begin{equation}
    A_3=0 \,, \quad F_X = 0\,, 
    % F=F(\phi)\,.
    \label{e:Hlimit}
\end{equation}
which restrict $F=F(\phi)$. It can be visualize in the Horndeski \cite{Horndeski:1974wa}, and GLPV notation \cite{Gleyzes:2014dya} as
\begin{equation*}
    F=G_4\,,\quad F_X=G_{4X}=0\,, \quad G_4=G_4(\phi)\,.
    %\label{e:Hlimit}
\end{equation*}
Therefore, viable Horndeski theory can be seen as a subclass of viable DHOST theory together with the conditions in Eq. \eqref{e:GLPVlimit} and Eq. \eqref{e:Hlimit}, which we refer as the Horndeski limit in our remaining paper. 

Graviton decay constraint: It is discussed in Refs. \cite{Creminelli:2018xsv,Creminelli:2019nok} that the gravitational waves may decay to dark energy perturbations. The constraint on the viable DHOST theory in Eq. \eqref{ac:dhost:gw} for avoiding this graviton decay is  
\begin{equation*}
 A_3 = 0\,,   
\end{equation*}
which fixes $A_5=0$.
We also investigate this $A_3 = 0$ subset of the DHOST theory, and named as A3eq0 in the rest of the paper (also see our discussion section \ref{sec:dis}), which was also been studied in \cite{Frusciante:2018tvu,Hirano:2019scf,Crisostomi:2019yfo}. The Horndeski theory can be seen as a subclass of A3eq0 by further setting $F_X=0$. 

From Eq.~\eqref{A_cg=1}, we find that
\begin{equation}
   A_5 =0\,, \quad \text{when}
   \begin{cases}
    \phantom{or } \frac{1}{2 F} =0 &\text{: unphysical} \\
    \text{or } 4{F_X}+X A_3 =0 & \text{: GLPV theory} \\ 
    \text{or } A_3 =0 & \text{: no graviton decay}   
   \end{cases} \nonumber 
\end{equation} 

Therefore, GLPV and ${\rm A3eq0}$ theories have the Horndeski theory in common, but are extended to a different sets of models. The relation of the functions of different theories, and their EFT parameters is discussed later in the Table \ref{tab:theory}. We use the ${\rm DHOST}$ theory in Eq.~\eqref{ac:dhost:gw} as our generalized framework in this paper.

%%%%%%%%%%%%%%%%%%%%%%%%%%%%%%%%%%%
\section{Characteristic parameters}
\label{sec:prams}
Modification of gravity can impact in both, background as well as perturbations. In this section, we parametrize the cosmological perturbations in the DHOST theory, which capture the modifications from GR in the linear perturbations. We adopt the low-energy single-field EFT of DE and MG parametrizations which describes a cosmological background evolution and the linear perturbations around it \cite{Gubitosi:2012hu, Bloomfield:2012ff,Langlois:2017dyl}. These minimal EFT parameters, $\alpha_M$, $\alpha_K$, $\alpha_B$, $\alpha_H$, and $\beta_1$ represent the observational deviation of a model in the ${\rm DHOST}$ theory from the $\Lambda$CDM model in the linear regime \cite{Bellini:2014fua,Langlois:2017mxy}. In this article, we are considering the detuning of the extrinsic curvature parameter, $\alpha_L = 0$ for the degeneracy class \cite{Langlois:2017mxy}. The excess tensor speed parameter is set to $\alpha_T=0$ since we consider the DHOST theory with $c_g=1$. It is worth mentioning that $\alpha_{K,B,M}$ parameters are shared with the Horndeski theory, but have the terms that only appear in the ${\rm DHOST}$ theory. As discussed in the introduction that our purpose here is to figure out the deviations of the ${\rm DHOST}$ theory from its largest subset passes through the screening mechanism named the Horndeski theory \cite{Kimura:2011dc,Narikawa:2013pjr,Koyama:2013paa,Kase:2013uja}. Therefore, our purpose here is to figure out the deviations from the Horndeski theory in the ${\rm DHOST}$ theory. In order to see the deviations, it is convenient to express those EFT parameters into two parts: For finding the deviation of the DHOST theory from the Horndeski theory, we split the EFT parameters into two following parts,
\begin{equation}
    \alpha_{M,K,B}^{\rm{DHOST}} = \alpha_{M,K,B}^{\rm Horn} + \alpha_{M,K,B}^{\rm res} \,,
\end{equation}
where $\alpha_{M,K,B}^{\rm Horn}$ characterizes the Horndeski theory, and  $\alpha_{M,K,B}^{\rm res}$ characterizes the deviations from the Horndeski theory, i.e., $\alpha_{M,K,B}^{\rm res} = 0$ gives the Horndeski limit in Eq.~\eqref{e:Hlimit}. We compute the EFT parameters of the DHOST theory in Appendix.~\ref{app:EFT} and the expressions are given below. The running of the effective Planck mass $M_*=\sqrt{2F}$ is given as
\begin{eqnarray}\label{nu_GW}
\alpha_M = \frac{1}{HF}\frac{d F}{dt} &=& \alpha^{\rm Horn}_M + \alpha^{\rm res}_M \,, \label{alphaM}
\end{eqnarray}
 where
\begin{align}
   &\alpha^{\rm Horn}_M \equiv \frac{\dot{\phi}F_\phi}{HF}\,,\label{aM_Horn} \\
    &\alpha^{\rm res}_M \equiv \frac{\dot{X}F_X}{HF}\,\label{aMHorn-res}. 
\end{align}
Here and hereafter, the dot in $\dot{\phi}$ denotes the time derivative with respect to the cosmic time. The effective Planck mass unchanges in time when $\alpha_M=0$. The parameter $\alpha_M$ becomes nonzero in general for modified gravity theories. The above expressions explain that $\alpha^{\rm res}_M$ has a similar structure for the GLPV, A3eq0, and DHOST theories. 

The kinetic braiding parameter or mixing of the kinetic terms of the scalar and metric is given by 
\begin{align}
\alpha^{\rm Horn}_B 
&\equiv \frac{\dot{\phi}(Q_X X + F_{\phi})}{2FH} \;, \label{aB_Horn}\\
\alpha^{\rm res}_B 
&\equiv \frac{1}{4FH}\left [ 
8HXF_X
+4\dot{\phi}XF_{\phi X}
-4\dot{\phi}\ddot{\phi} \left(3F_X + 2X F_{XX}  \right)
-\dot{\phi}\ddot{\phi}X(5A_3+2XA_{3X})\right ]
\,.\label{aB_res}
\end{align}

Notice that the third and fourth terms of $\alpha^{\rm res}_B$ vanish for the GLPV theory and while only the fourth term vanishes for the A3eq0 theory. Therefore, the expression of the $\alpha^{\rm res}_B=0$ is different for the GLPV, A3eq0, and DHOST theories.

The parameter $\alpha_K$, commonly appearing in the EFT parameters, is also computed in the ${\rm DHOST}_{c^2_g=1}$ theory and denotes the coefficient of the scalar perturbation. Since we use $\alpha_K$ only for assessing the stability conditions throughout this paper, we omit the specific expression of $\alpha_K$ here (see Appendix.~\ref{app:EFT} for the explicit form of $\alpha_K$). Beside the aforementioned $\alpha^{\rm res}_{M,K,B}$, two additional EFT parameters associated to the deviations of the DHOST theory from the Horndeski theory are $\alpha_H$ and $\beta_1$ expressed as 
\begin{align}
&\alpha_H = -\frac{2XF_X}{F}\,, \label{alphaH}\\
&\beta_1 = \frac{X(4F_X+A_3 X)}{4F}\,.\label{beta1}
\end{align} 
$\alpha_H$ denotes the additional appearance of higher derivative terms in GLPV theory \cite{Gleyzes:2014dya,Gleyzes:2014qga}. $\beta_1$ plays a similar role as kinetic-braiding parameter $\alpha_B$, being associated with the higher derivative terms in DHOST theory \cite{Langlois:2017dyl}. Notice that $\beta_1=0$ for the GLPV theory, and $\beta_1=XF_X/F$ for A3eq0 theory. Therefore, the GLPV, A3eq0 and DHOST theories traces different expressions for $\beta_1$, while the same expressions for $\alpha_H$.

The summary of the characteristic EFT parameters of the DHOST theory is displayed in Table~I. Note that setting $\beta_1=0$ in the A3eq0 theory also forces $\alpha_H$ to zero and becomes the Horndeski theory. It also confirms that both the GLPV and A3eq0 theories have the Horndeski in common (ref. to section \ref{sec:action}) and extended to different sets of EFT parameter space in the perturbation levels. In Sec.~\ref{sec:modeldist}, we propose a method to distinguish the theories within the DHOST theory via $\alpha_{M,B,H}$ and $\beta_1$. 

\begin{table*}[htb]
\begin{tabular}{||l||c|c|c||c|c|c|c|c|c||}
\hline\hline
\multirow{3}{*}{\backslashbox[18mm]{Theory \\ ${c_g=1}$ 
}{\phantom{.}}}
&\multicolumn{3}{c||}{arbitrary functions}
&\multicolumn{6}{c||}{linear parameters}
\\ \cline{2-10}
& \multirow{2}{*}{$F(\phi)$} & \multirow{2}{*}{$F_X$} & \multirow{2}{*}{$A_3$}
&\multicolumn{2}{c|}{$\alpha_M$}&\multicolumn{2}{c|}{$\alpha_B$} & \multirow{2}{*}{$\alpha_H$} & \multirow{2}{*}{$\beta_1$}
\\ \cline{5-8}
&&&& $\alpha_M^{\rm Horn}$&$\alpha^{\rm res}_M$ & $\alpha^{\rm Horn}_B$ &$\alpha^{\rm res}_B$&& \\ \hline \hline
Horneski$_{c_g=1}$ &\CheckmarkBold & 0 & 0& \CheckmarkBold & 0 & \CheckmarkBold & 0 & 0 &0 
\\ \hline
GLPV$_{c_g=1}$& \CheckmarkBold & \CheckmarkBold & 
$A_3 = -4F_X/X$
& \CheckmarkBold & \CheckmarkBold & \CheckmarkBold & \CheckmarkBold & \CheckmarkBold & 0  
\\ \hline
%%%%
${\rm A3eq0}_{c_g=1}$& \CheckmarkBold & \CheckmarkBold  & 0  & \CheckmarkBold & \CheckmarkBold & \CheckmarkBold & \CheckmarkBold & \CheckmarkBold & $\beta_1= - \alpha_H /2$ 
\\ \hline
%%%%
${\rm DHOST}_{c_g=1}$ & \CheckmarkBold & \CheckmarkBold & \CheckmarkBold & \CheckmarkBold & \CheckmarkBold & \CheckmarkBold & \CheckmarkBold & \CheckmarkBold  & \CheckmarkBold 
\\ \hline \hline
\end{tabular}
\caption{\label{tab:theory}Distinction of the theories with $c_g=1$ by EFT parameters and arbitrary functions.}
\end{table*}

%%%%%%%%%%%%%%%%%%%%%%%%%%%%%%%%
\section{Numerical formulation of DHOST theory}\label{sec:setup}

The characteristic behaviors of the aforementioned EFT parameters can be understood if one could find a cosmological solution of a scalar field, $\phi$, and gravitational perturbations of the full ${\rm DHOST}$ theory. Except for the exact solutions given in \cite{Crisostomi:2018bsp,Frusciante:2018tvu}, a cosmological solution of the full ${\rm DHOST}$ theory for general arbitrary functions in all redshifts regimes is unsolved yet. The full numerical solution has neither been studied nor been computationally cheap. When it comes to study only on the range of the observable variables or the EFT parameters, however, numerical optimizations would be found. The observationally viable Horndeski models were studied in model-independent way in \cite{Gleyzes:2017kpi,Arai:2017hxj, Nishizawa:2019rra}. Here, we apply the technique suggested in \cite{Arai:2017hxj,Nishizawa:2019rra} for the DHOST theory. First, we approximate $\phi$ as a function in time, and the arbitrary free functions, $P,Q, F, A_3$ by using Taylor series expansion without solving the background Friedmann equations of the system. Then we will check the stability and consistency with the observations of those solutions. 

%%%%%%%%%%%%%%%%%%%%%%%%%%%%%%%%%
\subsection{Approximation of the scalar field and arbitrary functions} \label{ssec:slow}

The challenge is to find the right parametrization for the evolution of the scalar field for which the Taylor series expansion would be pertinent in the late time Universe as well as in the early Universe up to the redshift $z \sim 1000$ of the cosmic microwave background (CMB) last scattering. 

A simple choice of the expansion argument of the scalar field $\phi$ is the inverse of the redshift. Though that expansion works well only in high redshifts, $z>1$, and diverges in smaller redshifts. An alternative possibility of the Taylor expansion parameter is the scale factor, $a$, which would work well for $a\le 1$. What the price of these parametrizations are that the time evolution of the scalar field is non-trivially related to the scale factor $a$ or $z$. From the physical point of view, it seems not natural to use $a$ or $z$ for the approximation of $\phi$. 

Then another possibility comes along with the expansion with the time variables, such as the look back (LB) time. The evolution of $\phi$ in terms of the cosmic LB time, $t_{\rm LB}(a) \equiv \int^1_a{d\tilde{a}/H(\tilde{a})\tilde{a}}$, works well in $z<1$, while the cosmic LB time quickly converges around $z=1$, making few features of the time evolution of $\phi$ \cite{Arai:2017hxj}. On the other hand, the conformal time is also the local time variable in each cosmological epoch. The region of convergence would also include the region $z\ge 1$. Therefore, if we want the time evolution of the scalar field which is valid for both regimes, the late-time (today) as well as the early Universe $(z \sim 1000)$, then the automatic choice for the expansion of the scalar field evolution is the look back conformal time, $\tau_{\rm LB}$, 
\begin{eqnarray}\label{tauLB}
\tau_{\rm LB} (a) = \int^1_a{\frac{d\tilde a}{H(\tilde a)\tilde a^2}}\,.
\end{eqnarray}

To connect the model-predictions to observations, we would first perform the Taylor expansion of the scalar field $\phi$ in the LB conformal time, and later express it and its time derivatives in terms of the scale factor, i.e., $\phi(a)$ and $\dot{\phi}(a)$. We assume that the scalar field changes slowly in time in comparison to the time scale of the cosmic expansion so that we can truncate the expansion at a finite order as 
\begin{align}\label{phi}
\phi(\tau_{\rm LB}) = M_\phi \left \{ b_0+b_1 H_0 \tau_{\rm LB} + \frac{b_2}{2} (H_0 \tau_{\rm LB})^2 +\frac{b_3}{6}(H_0 \tau_{\rm LB})^3\right \}\,,
\end{align}
where $M_\phi$ is the mass scale of $\phi$ at present. Notice that the range of $b_i$ is still arbitrary. Here we keep up to the third order in $\tau_{\rm LB}$ because we believe that even a relatively fast-evolving solution like the tracker solution \cite{Frusciante:2018tvu} is marginally included. The tracker solution is $\dot{\phi} \propto 1/H \propto t$, then $\phi \propto t^2$. On the other hand, in the matter-dominated Universe, since the conformal time and the cosmic time are related by $\tau^n \propto t^{n/3}$, the tracker solution ($n=6$) might apparently look excluded in our expansion. However, we are interested in the tracker solution after the late matter-dominated era ($z \leq 1$) where the difference between the cosmic time and the conformal time is not large (only a factor of ${\cal O} (1)$), the tracker solution is practically captured by our expansion at the order of $n=2$, even the scalings are different.

By using the formula Eq.~\eqref{tau_app}
(see Appendix \ref{app:taylor_expansion} for the detailed derivation), the evolution of $\phi(a)$ is rewritten as
\begin{align}
\hat{\phi} \equiv \phi(a)/\tilde{M}_\phi = c_0 + \sum^3_{i=1}{ c_i (1-a^{i/2})}\,.\label{e:phi:dim_less}
\end{align}
where $\tilde{M}_\phi$ is given in Eq.~\eqref{phi_ini}. We assign that the coefficients $c_i$ are utterly random in the range $[-1,1]$. Note that we are precisely sampling $c^{(n)}_\phi$ in Eq.~(37) of \cite{Nishizawa:2019rra}.

\begin{figure}[ht]
  \begin{center} 
    \includegraphics[clip,width=12.0cm]{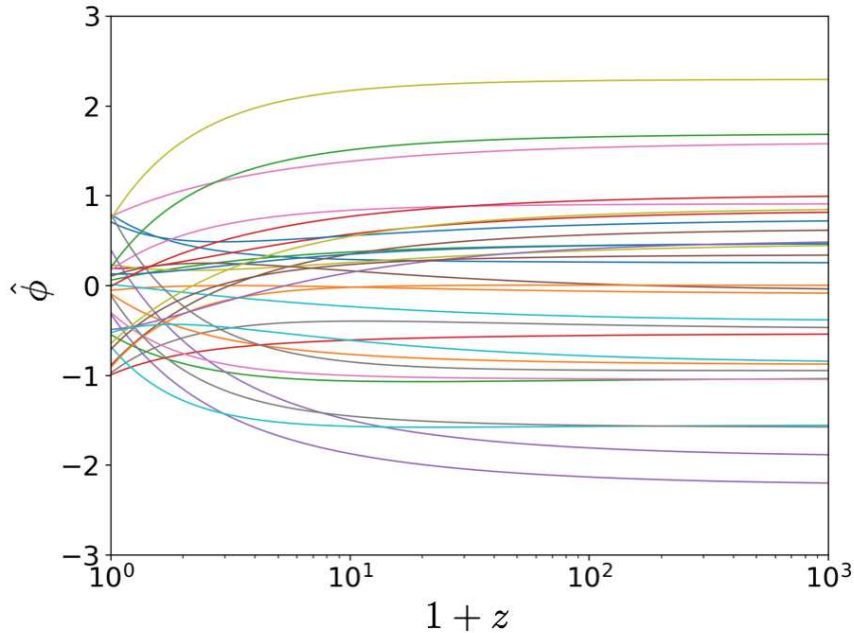}
    \caption{Time variation of $\phi$ with different $c_i (i=0,1,2,3)$ with the fixed $H_0$ and $\Omega_{\rm m0}$. $H_0$ and $\Omega_{\rm m0}$ are set to the best fit values by Planck 2015 \cite{Ade:2015xua}, $H_0 = 67.8\,{\rm km}\,{\rm s}^{-1}\,{\rm Mpc}^{-1}$ and $\Omega_{\rm m0} = 0.3080$.}
    \label{fig:phi_app}
  \end{center}
\end{figure}

Fig.~\ref{fig:phi_app} shows the evolution of $\phi(a)$ with respect to the redshift $1+z=a^{-1}$. We see that the scalar field evolves at low redshifts while converging to constant at high redshifts. Note that this result is consistent with the picture such that the scalar field slowly changes in time.

The dimensionless arbitrary functions as a function of time are
\begin{equation}\label{e:A:dim_less}
{\hat{\cal A}}^{(\rm app)}_i \equiv \frac{{\cal A}^{(\rm app)}_i(\phi,X)}{{\cal A}_i (\Lambda_P, \Lambda_Q)} = {a_i}+\sum_{\rho = \hat{\phi}, \hat{X}}a_{i ,\rho}\rho + \sum_{\rho, \sigma = \hat{\phi}, \hat{X}}\frac{a_{i,\rho \sigma}}{2}\rho \sigma + \sum_{\rho, \sigma, \lambda = \hat{\phi}, \hat{X}}\frac{a_{,i \rho \sigma \lambda}}{6}\rho \sigma \lambda \,, 
\end{equation}
where $\hat{X}\equiv -\dot{\phi}^2/{\tilde{M}_\phi}^2 H^2$. ${\cal A}^{(\rm app)}_i(\phi,X)$ with $i = 1, 2, 3, 4$ represent the DHOST theory functions, $P$, $Q$, $F$, or $A_3$, respectively. $H_0$ is the Hubble constant of today. $\tilde{M}_\phi$ and ${\cal A}_i (\Lambda_P, \Lambda_Q)$ are the normalization factors to make $\phi$ and ${\cal A}^{(\rm app)}_i(\phi,X)$ dimensionless,
\begin{equation}\label{calA_i}
{\cal A}_1 = \Lambda^4_P\,,\  {\cal A}_2 = \frac{\Lambda^4_P}{\Lambda^3_Q}\,, \ {\cal A}_3 = \frac{\Lambda^8_P}{\Lambda^6_Q}\,,\ 
{\cal A}_4 = \frac{1}{\Lambda^6_Q}\,,
\end{equation}
where $\Lambda_P \equiv (\tilde{M}_\phi H_0)^{1/2}$ and $\Lambda_Q \equiv (\tilde{M}_\phi H^2_0)^{1/3}$, respectively describing the dynamical energy scale of $\phi$ and the cut-off scale of non-linearity of $\phi$ at the present Hubble scale, $H_0$. Note that the cosmic acceleration realizes since ${\cal E}$ is at the order of the cosmic critical density, $M^2_{\rm pl}H_0^2$.

The above expressions are valid for the both, the late and early Universe. The model coefficients, $c_i$ ($i=0,1,2,3$) and $a_i$ ($i=1,2,3,4$), are the inputs in the numerical program, which are randomly chosen in the range of [-1,1]. This choice of the range is motivated by our normalizations in Eqs.~(\ref{e:phi:dim_less}) and (\ref{e:A:dim_less}). A particular set of values of $a_i$ $(i = 1,2,3,4)$ in Eq.~\eqref{e:A:dim_less} represents a model within the framework of the ${\rm DHOST}$ theory, and a set of $c_i$ in Eq.~\eqref{e:phi:dim_less} represents the time evolution of the scalar field in that model. Given the expressions of $\phi(a)$ and ${\cal A}^{(\rm app)}_i (a)$, we would able to evaluate all EFT parameters, $\alpha_{M,B,H}$, and $\beta_1$, mentioned in the previous section by using Eqs.~ \eqref{e:phi:dim_less} and \eqref{e:A:dim_less}. 

These approximated scalar field evolutions do not guarantee that they all will satisfy the equations of motion of the model. As a prescription for being consistent, we will filter models by the conditions of i) theoretical stability and ii) observational constraint explained in the next subsection.

%%%%%%%%%%%%%%%%%%%%%%%%%%%%%%%%%%%%%%%%%
\subsection{Filtering through the consistency and stability conditions}\label{ssec:solver}

We check the following consistency and stability conditions at redshifts, $z=0$, 0.1, 0.5, 1.0, 1.5, and 2.0, where the constraints on the Hubble parameter exist \cite{Farooq:2016zwm}. In the following approximations, we use the Hubble expansion rate of the $\Lambda {\rm CDM}$ model,

\begin{align}
    H_{\Lambda {\rm CDM}} = H_0 \sqrt{\Omega_{\rm m0}a^{-3}+1-\Omega_{\rm m0}}\,,\label{HLCDM}
\end{align}
with $H_0 = 67.8\,{\rm km}\,{\rm s}^{-1}\,{\rm Mpc}^{-1}$ and $\Omega_{\rm m0}=0.3080$ from \cite{Ade:2015xua}, which is the same as in Appendix.~\ref{app:taylor_expansion}.

 (i) {\bf Consistency conditions:} In the previous section, we arbitrarily produced the numerical solution of $\phi$ without solving the Friedmann equation. Therefore, we will filter only the models which can consistently produce the Hubble parameter, $H$, and its time variation, $\dot{H} $, within the observational error, $20\%$ deviation from the $\Lambda \rm CDM$ model (Table~I of \cite{Farooq:2016zwm}).
 
 We substitute $H_{\Lambda \rm CDM}$ and $\phi(\tau_{\rm LB})$ in the right-hand sides of Eqs.~\eqref{F1} and \eqref{F2} which give $H_{\rm DHOST}$ and $\dot{H}_{\rm DHOST}$. Then we check two following consistency filters for the Hubble parameter(FH) and the derivative of the Hubble parameter (FdH),
 \begin{eqnarray}
 {\rm FH}: \biggl | 1- H_{\rm DHOST}/H_{\rm \Lambda CDM} \biggr | < 20\% \,, \label{FH}\\
 {\rm FdH}: \biggl | 1- \dot{H}_{\rm DHOST}/\dot{H}_{\rm \Lambda CDM} \biggr | < 20\%\,. \label{FdH} 
 \end{eqnarray}
 These consistency conditions guarantee the evolution of $\phi(\tau_{\rm LB})$ within the observational ranges of the Hubble parameter and its changes. In particular, we verified that the filtering condition on $\dot{H} $ effectively excludes relatively-fast evolving models.

(ii) {\bf Stability conditions:} For ensuring the linear scalar and tensor perturbations are free from ghost and gradient instabilities, we pass through the stability conditions \cite{Crisostomi:2018bsp}, 
\begin{align}
&A_{\tilde{\zeta}}+\frac{\rho_{m}+p_{m}}{M^2_*H^2_{\Lambda {\rm CDM}}} \frac{3 \beta_1(2+3c^2_{\rm m}\beta_1)}{(1+\alpha_B-\dot{\beta}_1/H_{\Lambda {\rm CDM}})^2}>0\,,\ \cr 
&B_{\tilde{\zeta}}+\frac{\rho_{m}+p_{m}}{M^2_*H^2_{\Lambda {\rm CDM}}}\left( \frac{1+\alpha_H+\beta_1}{1+\alpha_B-\dot{\beta}_1/H_{\Lambda {\rm CDM}}}\right)^2 < 0\,,\ M^2_*>0 \,. \label{stab}
\end{align} 
All the aforementioned quantities are derived and defined in the Appendix \ref{app:EFT} (see Eqs.~\eqref{Azeta_wo_m} and \eqref{Bzeta_wo_m}). Please note that the inclusion of matter changes the stability conditions, since matter itself may introduce the instability\cite{Kase:2014cwa,Gleyzes:2014qga,DeFelice:2016ucp,Frusciante:2018vht, Ganz:2018mqi}. Linear stability may also depend on the chosen basis of scalar perturbations and particularly on nonzero $\alpha_H$ and $\beta_1$. Detailed discussion is in Appendix~\ref{app:EFT}.

\section{Discriminating theories via the distributions of characteristic parameters}\label{sec:modeldist}

In this section, we will demonstrate the correlations among the characteristic parameters; $\alpha_M$, $\alpha_B$, $\alpha_H$, and $\beta_1$, introduced in Sec.~\ref{sec:prams}, and present the model distributions in the parameter space as a function of redshifts $z$ by using the numerical techniques explained in the previous section \ref{sec:setup}. The model distribution is shown for each subgroup of the ${\rm DHOST}$ theory summarized in Table~\ref{tab:theory} and is interpreted based on the order-of-magnitude estimation. 

\subsection{Time evolutions of characteristic parameter distributions}\label{ssec:zev_charac}
 
\subsubsection{Distribution of $\alpha_M$}
 
\begin{figure}[htb]
  \begin{center} 
    \includegraphics[clip,width=12.0cm]{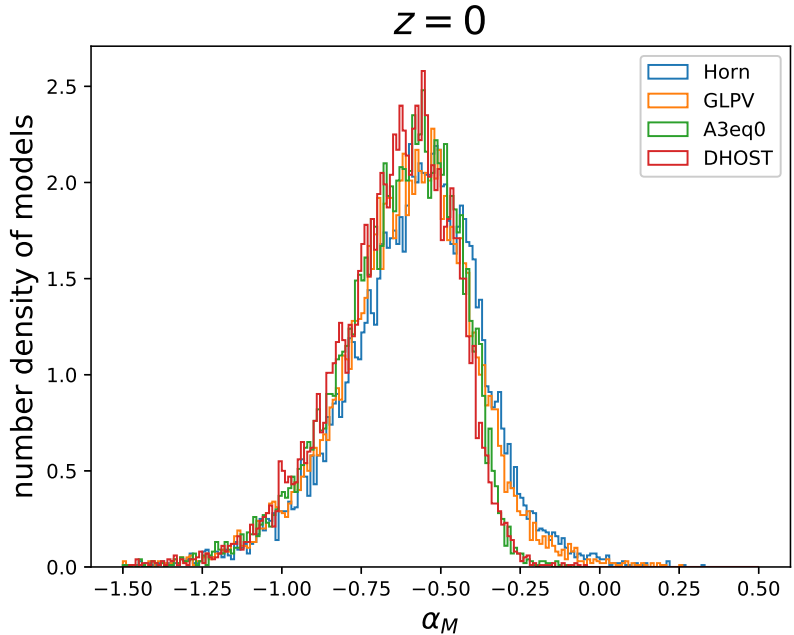}
    \caption{The model distribution in $\alpha_M$ at $z=0$ with the bin size $\Delta \alpha_M=0.01$. 
    The solid color lines represent Horndeski (blue), GLPV (orange), A3eq0 (green), and ${\rm DHOST}$ (red), respectively. This is the enlarged version of the top left panel in Fig.~\ref{fig:aM_zev}.} 
    \label{fig:aM_z0}
  \end{center}
\end{figure}

\begin{figure}[htb]
  \begin{center} 
    \includegraphics[clip,width=16.0cm]{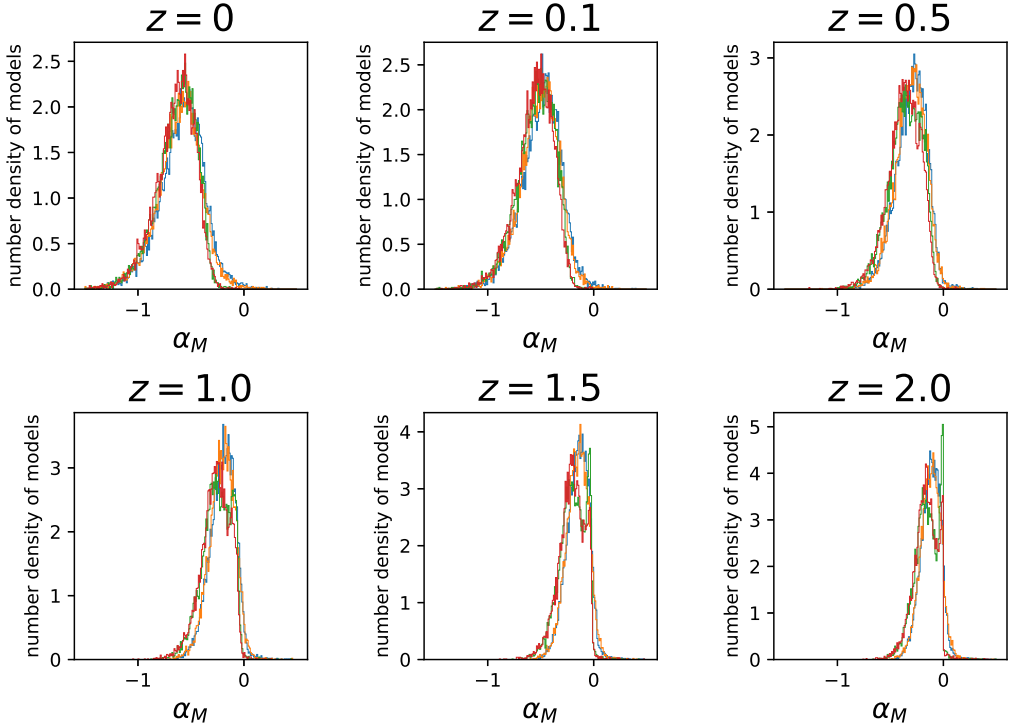}
    \caption{The model distribution in $\alpha_M$ at different redshift from $z=2.0$ to $z=0$ with the bin size $\Delta \alpha_M=0.01$.  
   The colors are the same as in Fig.~\ref{fig:aM_z0}}
    \label{fig:aM_zev}
  \end{center}
\end{figure}

By using the expansions of $\hat{\phi}$ and $\hat{\cal{A}}_i^{({\rm app})}$ in Eqs.~\eqref{e:phi:dim_less} and \eqref{e:A:dim_less}, we get the order of the $\alpha_M$ from Eqs.~\eqref{aMHorn-res} and \eqref{nu_GW}, 
\begin{align}
\alpha^{\rm Horn}_M =& {\cal O}(|\hat{X}|^{1/2})\,,\quad  \alpha^{\rm res}_M = {\cal O}(|\hat{X}|)\,, \\
\alpha_M &= \alpha^{\rm Horn}_M \left( 1 + {\cal O}(|\hat{X}|^{1/2}) \right)\,\label{aM_app}\\
&\simeq \alpha^{\rm Horn}_M \qquad \hbox{since} \quad  |\hat{X}| \ll 1.  
\end{align}
Since the order of $\alpha_M$ of the ${\rm DHOST}$ theory is the same as the Horndeski theory, $\alpha_M$ parameter is almost identical. Indeed, the indistinguishability of $\alpha_M$ is confirmed from the distribution of $\alpha_M$ at all different redshifts, $z=0\,,0.1\,,0.5\,, 1\,, 1.5\,, 2.0$ in the Figs.~\ref{fig:aM_z0} and \ref{fig:aM_zev}.

Figures~\ref{fig:aM_z0} and \ref{fig:aM_zev} show that $\alpha_M$ has a peak around $\alpha_M \sim -0.5$ at all redshifts. The negative value can be intuitively interpreted from the energy balance of the Friedmann equation in \eqref{E}.
The energy density of ${\rm DHOST}$ theory from Eq.~\eqref{F1}
is
\begin{align}
{\cal E} &= V_{\rm eff} + {\cal O }(|\hat{X}|)M^2_{\rm pl}H^2\,,\\
\hbox{with}\quad  V_{\rm eff} &\equiv V(\phi)-3M^2_*H^2\alpha^{\rm Horn}_M\,.
    \label{Veff}
\end{align}
The potential $V(\phi)$ is the sum over the $\phi$ dependence terms in $\cal E$. By inserting the above approximated $\cal E$ into Eq.~\eqref{E}, the Friedmann equation becomes
\begin{align}
    1 = \frac{V_{\rm eff}}{3M^2_*H^2} + \frac{\rho_m}{3M^2_*H^2}+ {\cal O}(|\hat{X}|)\,. \label{F1_app}
\end{align}
The matter density is negligible during the cosmic acceleration, resulting in $V_{\rm eff} \sim 3M^2_*H^2$. Because the models are drawn by random coefficients, all terms in $V_{\rm eff}$ are equally significant at the same order in $\hat{X}$, ending up with $-3M^2_*H^2\alpha^{\rm Horn}_M \sim V \sim 0.5V_{\rm eff}>0$. The negative value of $\alpha_M$ has already been encountered in our previous investigation of the Horndeski theory~\cite{Nishizawa:2019rra} under the assumption of $|\hat{X}| \ll 1$. In summary, $\alpha_M$ does not tell the difference between ${\rm DHOST}$ theory from the Horndeski theory in the observations.

\begin{figure}[htb]
  \begin{center} 
    \includegraphics[clip,width=16.0cm]{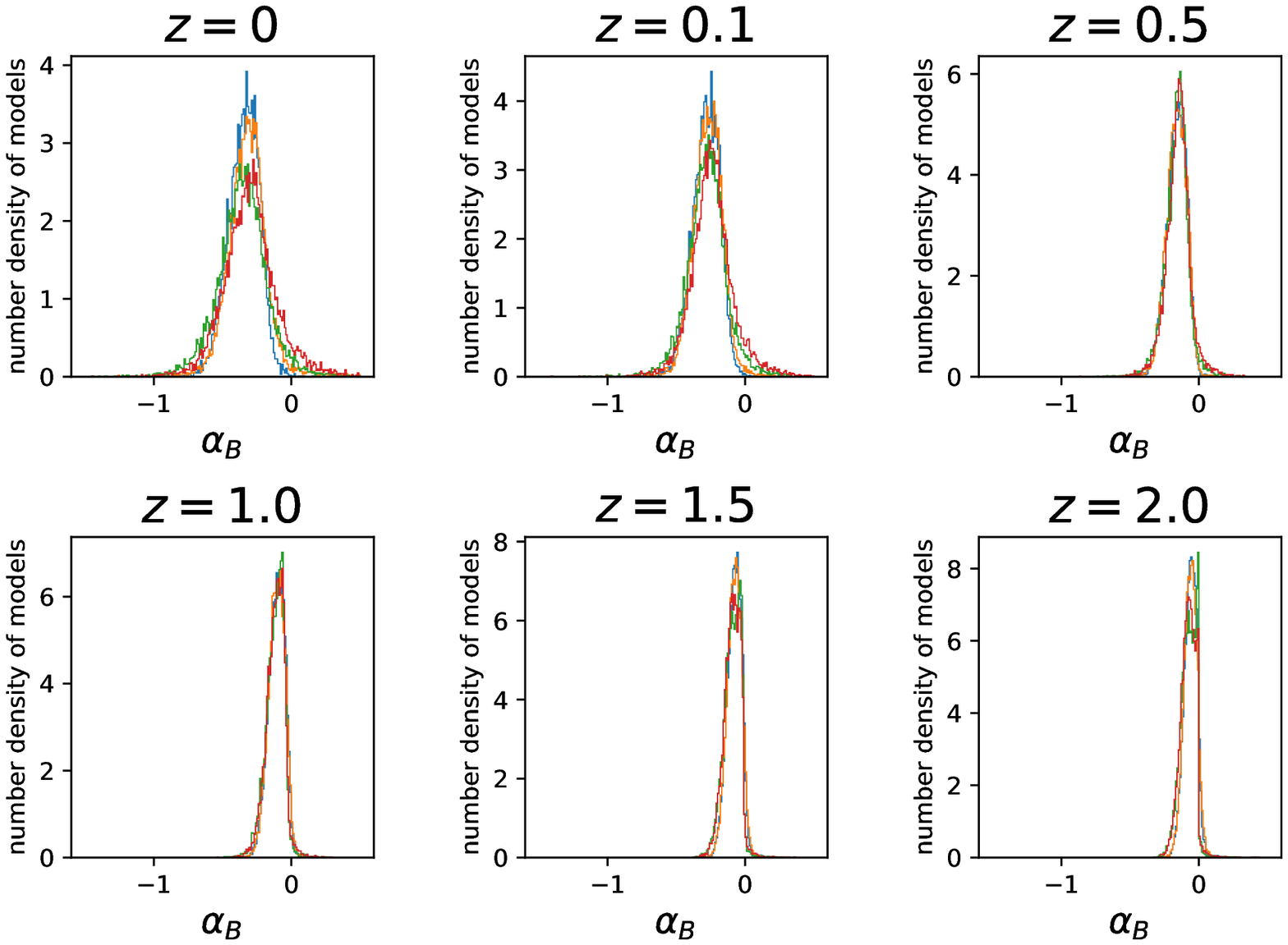}
   \caption{The model distribution in $\alpha_B$ with the bin size $\Delta \alpha_B=0.01$. 
   The colors are the same as in Fig.~\ref{fig:aM_z0}}
    \label{fig:aB_zev}
  \end{center}
\end{figure}

\subsubsection{Distribution of $\alpha_B$}
In contrast to $\alpha_M$, Fig.~\ref{fig:aB_zev} shows that the theories slightly deviates from Horndeski theory around the right tail of the distribution at $z=0$ and $z=0.1$, while indistinguishable at other parts of the distribution or above $z=0.5$. For all the theories, the locations of the peaks of all the distributions at all the redshifts are almost identical, and biased toward the negative side. In more detail, the distribution for the GLPV theory in orange is almost identical to that of the Horndeski theory. The reason of the above characteristics are understood from Eq.~\eqref{aB_Horn} as follows. The first term of $\alpha^{\rm Horn}_B$, $\dot{\phi}Q_X X/2FH$, is of the order of ${\cal O}(|\hat{X}|^{3/2})$ and the second term $\dot{\phi}F_\phi/2FH$ is exactly the same as $\alpha^{\rm Horn}_M/2$ and is of the order of ${\cal O}(|\hat{X}|^{1/2})$. One can derive the order of the $\alpha_B$ from Eqs.~\eqref{aB_Horn} and \eqref{aB_res},
\begin{align}
\alpha^{\rm Horn}_B =& {\cal O}(|\hat{X}|^{1/2})\,,\quad  \alpha^{\rm res}_B = {\cal O}(|\hat{X}|)\,, \label{order:aB:Horn_res}\\
\alpha_B &= \frac{\alpha^{\rm Horn}_M}{2}\left(1 + {\cal O}(|\hat{X}|^{1/2}) \right)\,,\label{aB-aM_app}
\end{align}
The leading term of Eq.~(\ref{aB-aM_app}) is $\alpha^{\rm Horn}_M$ , which is negative. Therefore, $\alpha_B$ is biased to the negative values in Fig.~\ref{fig:aB_zev}. The difference in $\alpha_B$ arises from the second term in Eq.~\eqref{aB-aM_app}, which is of the order of ${\cal O}(|\hat{X}|^{1/2}$). Earlier, we saw that the theories are hardly distinguishable in $\alpha_M$. Therefore, the locations of the distribution peaks in $\alpha_B$ are almost identical to $\alpha_M/2$.

$\alpha_B$ for the GLPV theory is well similar to that for the Horndeski theory. This is due to the hierarchy in the order of ${\cal O}(|\hat{X}|)$ is different from A3eq0 and DHOST theory. 
Recall $A_3X^2/F = {\cal O}(|\hat{X}|^2)$ for all the theories we consider, leading $\dot{\phi}\ddot{\phi}X(5A_3+2XA_{3X})/4FH = {\cal O}(|\hat{X}|^2)$.
Since $4F_X+A_3X=0$ is necessary for the GLPV theory and $\ddot{\phi} \sim H\dot{\phi}$, we obtain from Eq.~\eqref{aB_res} $\alpha^{\rm res}_B = {\cal O}(|\hat{X}|^2)$. As a result, $\alpha_B$ for the GLPV theory, i.e., $\alpha^{\rm GLPV}_B$ is peculiarly expressed by
\begin{align}
\alpha^{\rm GLPV}_B = 
\frac{\alpha^{\rm Horn}_M}{2}\left(1 + {\cal O}(|\hat{X}|^{3/2}) \right)\,.\label{aB-aM_app_GLPV}
\end{align}
Therefore, the GLPV theory is little ditinguished from the Horndeski theory.

%\sa{We interpret the \st{broader} distributions of the A3eq0 theory and the ${\rm DHOST}$ theory with the help of Eq.~(\ref{aB-aM_app}). The distributions of the two theories look like the superposition of the two components: the principal component is, making the peak of the distribution in the Horndeski theory, and the random component around the peak. \st{It is worth noting that the subleading contribution of $\alpha^{\rm res}_B$ in $\alpha_B$ is quantitatively larger than $\alpha^{\rm res}_M$ in $\alpha_M$ because the dimensionless coefficients multiplied by the terms $HXF_X$ and $\dot{\phi}\ddot{\phi}F_X$ in $\alpha^{\rm res}_B$ are relatively larger.}}\SA{I deleted the last sentence which is not obvious any more.}

The variance of $\alpha_M$ and $\alpha_B$ decrease as a redshift increases, because the time evolution of $\phi$ is slower at higher redshifts where matter starts to dominate, i.e.,~$|\hat{X}| \propto \dot{\phi}^2/H^2 \propto H^{-2/3}$, and the magnitudes of $\alpha_M$ and $\alpha_B$ are roughly given by $\alpha_M = {\cal O}(|\hat{X}|^{1/2})$ and $\alpha_B = {\cal O}(|\hat{X}|^{1/2})$.

\subsubsection{Distribution of extended Horndeski parameters, $\alpha_H$ and $\beta_1$}

\begin{figure}[t]
  \begin{center} 
    \includegraphics[clip,width=16.0cm]{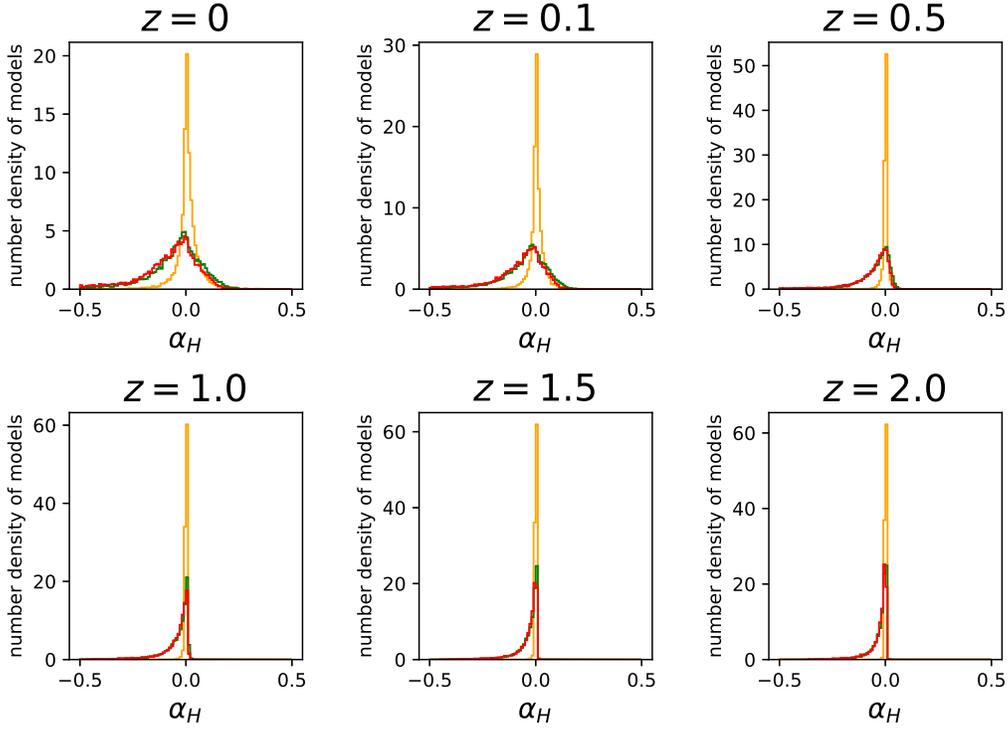}
   \caption{The model distribution in $\alpha_H$ with the bin size $\Delta \alpha_H=0.01$. 
   The colors are the same as in Fig.~\ref{fig:aM_z0}. We do not plot the Horndeski theory in which $\alpha_H=0$.}
    \label{fig:aH_zev}
  \end{center}
\end{figure}

\begin{figure}[t]
  \begin{center} 
    \includegraphics[clip,width=16.0cm]{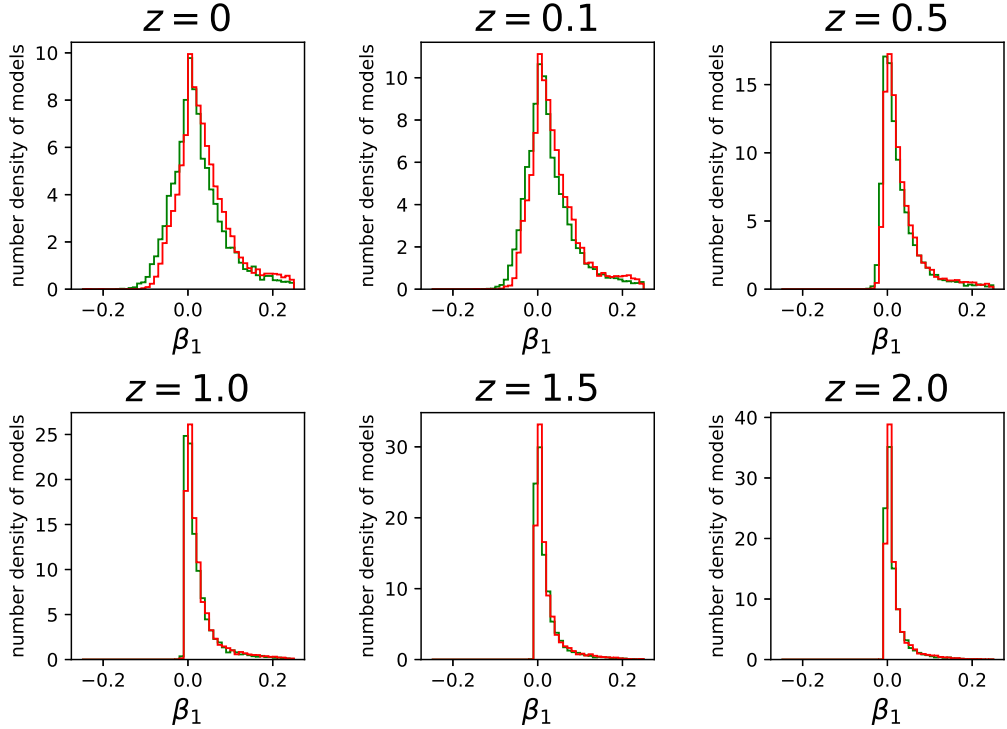}
    \caption{The model distribution in $\beta_1$ with the bin size $\Delta \beta_1=0.01$. The colors are the same as in Fig.~\ref{fig:aM_z0}. We do not plot the Horndeski theory and the GLPV theory in which $\beta_1=0$.}
    \label{fig:b1_zev}
  \end{center}
\end{figure}

The models distributions in $\alpha_H$ and $\beta_1$ are shown in Figs.~\ref{fig:aH_zev} and \ref{fig:b1_zev}, respectively. At first glance, $\alpha_H$ is evenly scattered around zero for the three plotted theories. The A3eq0 and the ${\rm DHOST}$ are distributed almost identically in $\alpha_H$. In the GLPV theory, the models are highly concentrated around zero due to the condition for the GLPV theory, i.e.,~$4F_X+A_3X=0$. Using this relation into the definition of $\alpha_H$ in Eq.~(\ref{alphaH}), we have $\alpha_H=-2XF_X/F=A_3X^2/2F={\cal O}(|\hat{X}|^2)$. Consequently, the models in the GLPV theory are peaked sharply at $\alpha_H = 0$. For the other theories, A3eq0 and ${\rm DHOST}$, $\alpha_H = {\cal O}(|\hat{X}|)$ still keeps the distributions peaked at $\alpha_H = 0$, in contrast to $\alpha_M = {\cal O}(|\hat{X}|^{1/2})$ and $\alpha_B = {\cal O}(|\hat{X}|^{1/2})$, merely because of a higher order contribution.

The other parameter $\beta_1$ is shown in Fig.~\ref{fig:b1_zev} and has a subtle difference in the distributions between A3eq0 and ${\rm DHOST}$. This is because the function which discriminate A3eq0 and ${\rm DHOST}$ is $A_3$, whose term is always subleading in $\beta_1$, e.g.,~$A_3X^2/F = {\cal O}(|\hat{X}|^2)$. From Eqs.~(\ref{alphaH}) and (\ref{beta1}), we obtain the relation
\begin{align}
\alpha_H = -2\beta_1 + {\cal O}(|\hat{X}|^2)\,. \label{aH2b1_app}    
\end{align}
After all, $\alpha_H$ and $\beta_1$ are dependent up to the order of ${\cal O}(|\hat{X}|)$. The difference begins to arise at the orders higher than ${\cal O}(|\hat{X}|)$. 

We summarize the following remarks. For all the theories in Table~\ref{tab:theory}$, \alpha_M$ hardly tells the differences among the theories via Eq.~(\ref{aM_app}).
In contrast, $\alpha_B$ at the tail of the distribution gives marginal discrimination among the theories. This is because $\alpha_B^{\rm res}$ contains multiple additional terms. $\alpha_H$ is generally ${\cal O}(|\hat{X}|)$, and $\alpha_H$ and $\beta_1$ always correlates via Eq.~(\ref{aH2b1_app}). The terms associated with $A_3$ are subdominant throughout the features of $\alpha_M$, $\alpha_B$, $\alpha_H$, and $\beta_1$. Interestingly, we find specific features in the GLPV theory, $\alpha_B \approx \alpha^{\rm Horn}_B$ and $\alpha_H \approx 0$. These state that the condition $4F_X+A_3X=0$ for the GLPV theory selects out a fine-tuned theory from the ${\rm DHOST}$ theory as a model for the cosmic acceleration.

\subsection{Correlations between characteristic parameters}\label{ssec:corr_params}

\begin{figure}[ht]
\begin{center} 
    \includegraphics[width=8cm]{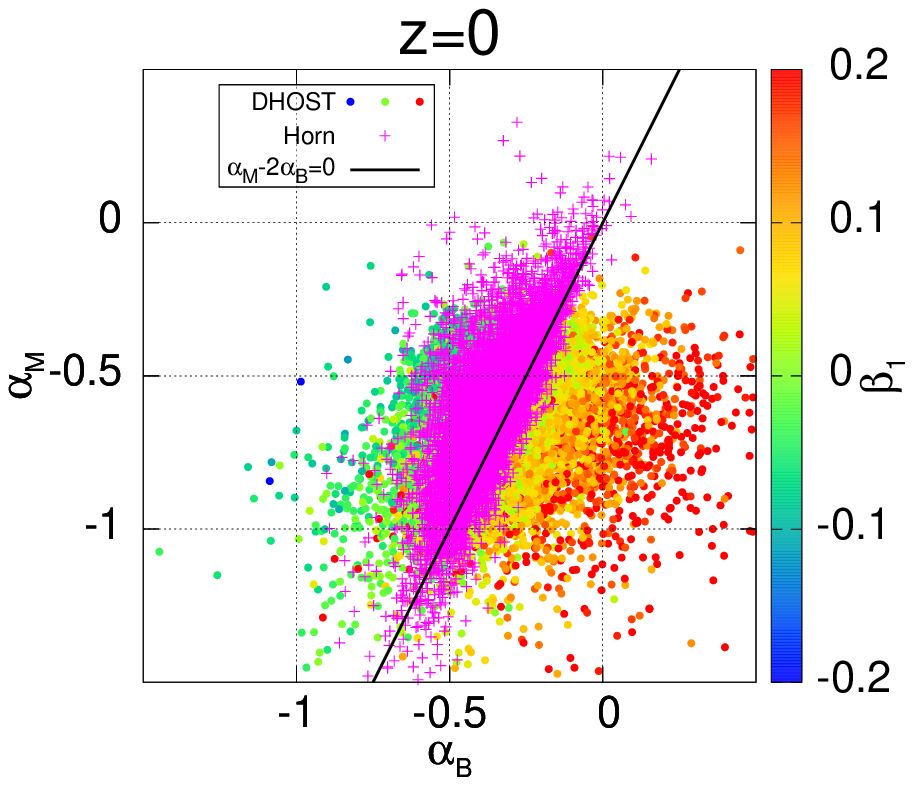} 
    \includegraphics[width=8cm]{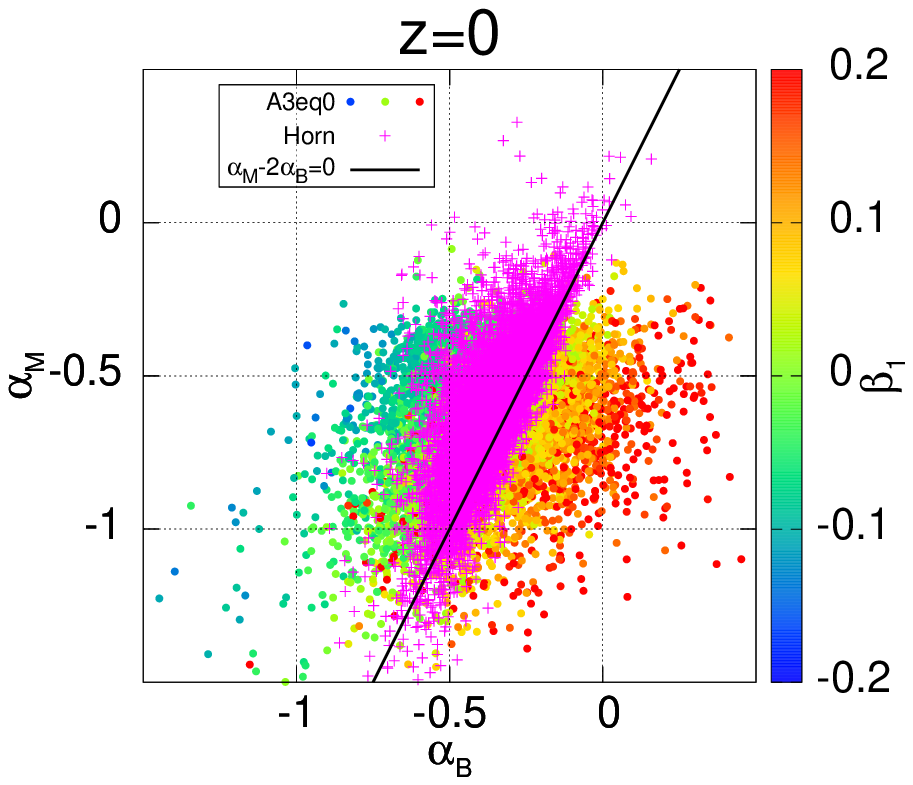} 
    \caption{Correlations among  $\alpha_M$,$\alpha_B$, and $\beta_1$. The panels show the distribution of models in the DHOST theory (left) and the A3eq0 theory (right), respectively. The cross points in magenta show the distribution of the Horndeski theory, which is overlapped partially with the distribution of the DHOST and A3eq0 theories.} 
   \label{fig:aM_aB_b1}
\end{center}
\end{figure}

We further investigate the differences among the theories via correlations among the four characteristic parameters, $\alpha_M$, $\alpha_B$, $\alpha_H$, and $\beta_1$. Since $\alpha_H$ is approximated by $-2\beta_1$ with the difference at the order of $|\hat{X}^2|$, we study the correlations among three parameters, $\alpha_M$, $\alpha_B$, and $\beta_1$. 

Figure~\ref{fig:aM_aB_b1} shows the model distributions in three dimensional parameter space composed of $\alpha_M$, $\alpha_B$, and $\beta_1$ in the ${\rm DHOST}$ and A3eq0 theories. At first sight, we confirm that the GR limit, i.e.,~$\alpha_M = 0 = \alpha_B$, is included in the Horndeski theory. In more detail, in the left panel, we confirm that the distributions of $\alpha_M$ and $\alpha_B$ for different values of $\beta_1$ in different colors are stretched along the line whose inclination is approximately 2 and clearly form the layers in parallel to a black line. In the right panel, the model distributions in the A3eq0 theory are shown. The features discussed above on the left panel also hold on the right except for that the distribution is slightly biased to smaller $\alpha_B$ and smaller $\beta_1$. 

One can find the following relations by expanding the analytic forms of the $\alpha_M$, $\alpha_B$, and $\beta_1$ given in Eqs.~(\ref{aMHorn-res}),~\eqref{aB_Horn}, and (\ref{beta1}) up to the leading and next-to-leading orders in $\hat{X}$.
\begin{align}
&{\cal O}(|\hat{X}|^{1/2}):\ \alpha^{\rm Horn}_M = 2\alpha^{{\rm Horn}}_B \,,\label{aMaB_sqrthatX}\\
&{\cal O}(|\hat{X}|):\ \alpha^{\rm res}_M
= \frac{2}{3}\alpha^{\rm res}_B - \frac{4}{3}\beta_1\,.
\label{aMaB_hatX} 
\end{align}
The inclination of the black line and the colored layers are originated from Eq.~(\ref{aMaB_sqrthatX}), which is the leading order relation. The inclination of the stretched distributions coincides with the coefficient in Eq.~(\ref{aMaB_hatX}), which gives deviation from the Horndeski theory. The continuous change of the color is characterized by the second term in Eq.~(\ref{aMaB_hatX}) and shows that
the ${\rm DHOST}$ and A3eq0 theories indeed deviate from the Horndeski theory (away from the black line). The domain of $\alpha_M$ and $\alpha_B$ with $\beta_1>0$ is significantly distinguishable. Discriminating the GLPV theories from the Horndeski theory is difficult, since $\alpha_B \approx \alpha^{\rm Horn}_B$, $\alpha_H \approx 0$, and $\beta_1=0$. 
\begin{figure}[htb]
  \begin{center} 
    \includegraphics[clip,width=12.0cm]{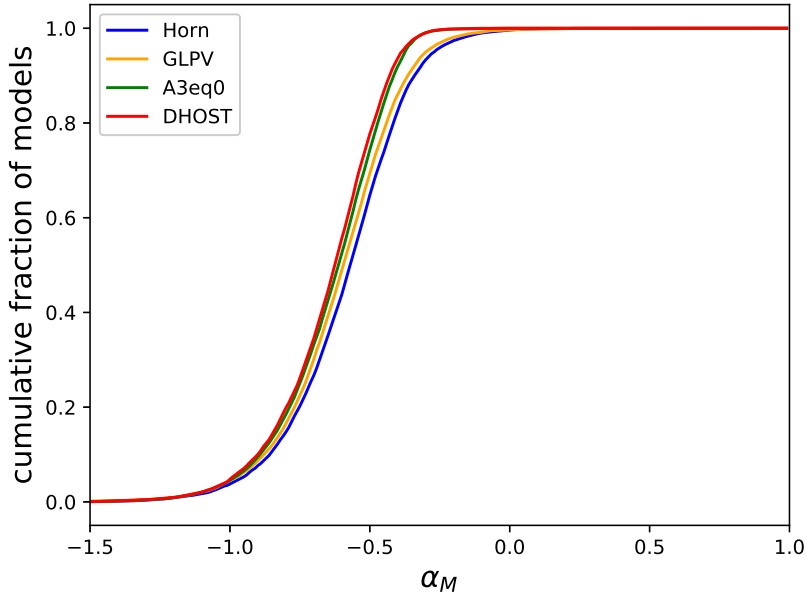}
   \caption{The cumulative fraction of models as a function of $\alpha_M$.}
    \label{fig:aM_cum}
  \end{center}
\end{figure}

\begin{figure}[htb]
  \begin{center} 
    \includegraphics[clip,width=12.0cm]{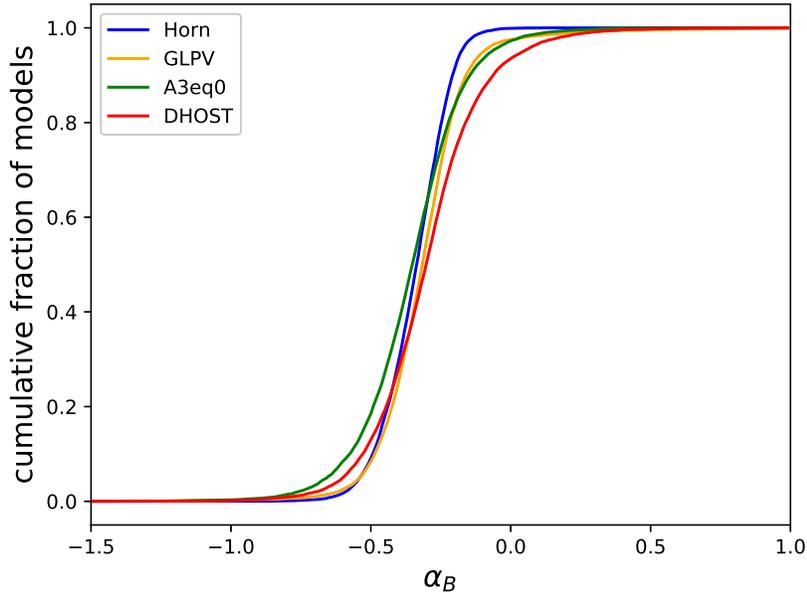}
   \caption{The cumulative fraction of models as a function of $\alpha_B$.}
    \label{fig:aB_cum}
  \end{center}
\end{figure}

Figures~\ref{fig:aM_cum} and ~\ref{fig:aB_cum} show the cumulative fraction of models as a function of $\alpha_M$ or $\alpha_B$ for each theory, respectively. In Fig.~\ref{fig:aM_cum}, more than $90\%$ of the models for all the theories are distributed in $-1 < \alpha_M < -0.3$, and there are few distinguished features in the shape of the lines for the corresponding theories. On the other hand, in Fig.~\ref{fig:aB_cum}, we notice that the model distributions in $\alpha_B<-0.9$ or $\alpha_B > 0.3$ is almost similar among the four theories, but in the intermediate range, the distribution of the theories slightly deviates with each other.

Theoretically, the difference of the Horndeski theory from the rest of the theories in $\alpha_B$ is characterized by $\alpha^{\rm res}_B$. By using the estimation in Eqs.~\eqref{aM_app},~\eqref{aMaB_sqrthatX} and ~\eqref{aMaB_hatX} and considering the indistinguishability of the theories in $\alpha_M$, we find that $\alpha_B-\alpha_M/2$ is the main component that discriminates the Horndeski theory from GLPV, A3eq0, and DHOST theory. Since Eqs.~\eqref{aH2b1_app} - \eqref{aMaB_hatX} reduce the six parameters $(\alpha_M,\alpha^{\rm Horn}_M,\alpha_B,\alpha^{\rm Horn}_B,\alpha_H,\beta_1)$ into three parameters, we conclude that $(\alpha_M,\alpha_B-\alpha_M/2,\beta_1)$ is a useful set of parameters to discriminate the DHOST theory from the Horndeski theory.

\begin{comment}
\begin{table*}[htb]
\begin{tabular}{lcccccl}
\hline
$\alpha_M (z=0)$ &\quad Horndeski \quad & \quad GLPV \quad & \quad A3eq0 \quad & \quad DHOST \quad\\ \hline \hline
$\alpha_M>-1$ & \quad 97 \quad & \quad 96 \quad & \quad 96 \quad & \quad 98 \quad \\
$\alpha_M>-0.1$ & \quad  $<1$ \quad & \quad $<1$ \quad & \quad $<1$ \quad & \quad $<1$ \quad \\
$\alpha_M>-0.01$ & \quad $<1$  \quad & \quad $<1$ \quad & \quad $<1$ \quad & \quad $<1$ \quad \\
\hline \hline
\end{tabular}
\caption{The percentage of the number of the models in the domains of $\alpha_M$. $\alpha_M$ is computed at $z=0$. The number of the models is fixed $10^4$ for all the theories. The number fluctuated by 1\% is not shown.}
\label{tab:aM_domain}
\end{table*}
\end{comment}

\begin{comment}
\begin{table*}[htb]
\begin{tabular}{lcccccl}
\hline
$\alpha_B (z=0)$ & \quad Horndeski \quad & \quad GLPV \quad & \quad A3eq0 \quad & \quad DHOST \quad\\ \hline \hline
$\alpha_B>-1$ & \quad 99 \quad & \quad 99 \quad & \quad 96 \quad & \quad 99 \quad\\
$\alpha_B>-0.1$ & \quad $<1$ \quad & \quad 4 \quad & \quad 6 \quad & \quad 19 \quad \\
$\alpha_B>-0.01$ & \quad $<1$ \quad & \quad 2 \quad & \quad 2 \quad & \quad 11 \quad\\
\hline \hline
\end{tabular}
\caption{The percentage of the number of the models in the domains of $\alpha_B$. $\alpha_B$ is computed at $z=0$. The number of the models is fixed $10^4$ for all the theories.The number fluctuated by 1\% is not shown.} 
\label{tab:aB_domain}
\end{table*}
\end{comment}

\subsection{Time evolutions of the principal parameters}\label{ssec:zev_aM_aB05aM_b1}

Here we demonstrate the time evolution of the principal set of parameters ($\alpha_M$, $\alpha_B-\alpha_M/2$, $\beta_1$) in the filtered Horndeski and DHOST theories. We focus on continuous
changes of the parameters of each model in the redshift range of $0 \le z \le 5$, where each parameter starts to deviate significantly from zero and to accelerate the cosmic expansion.

\begin{figure}[h]
\begin{center}
    \includegraphics[width=8cm]{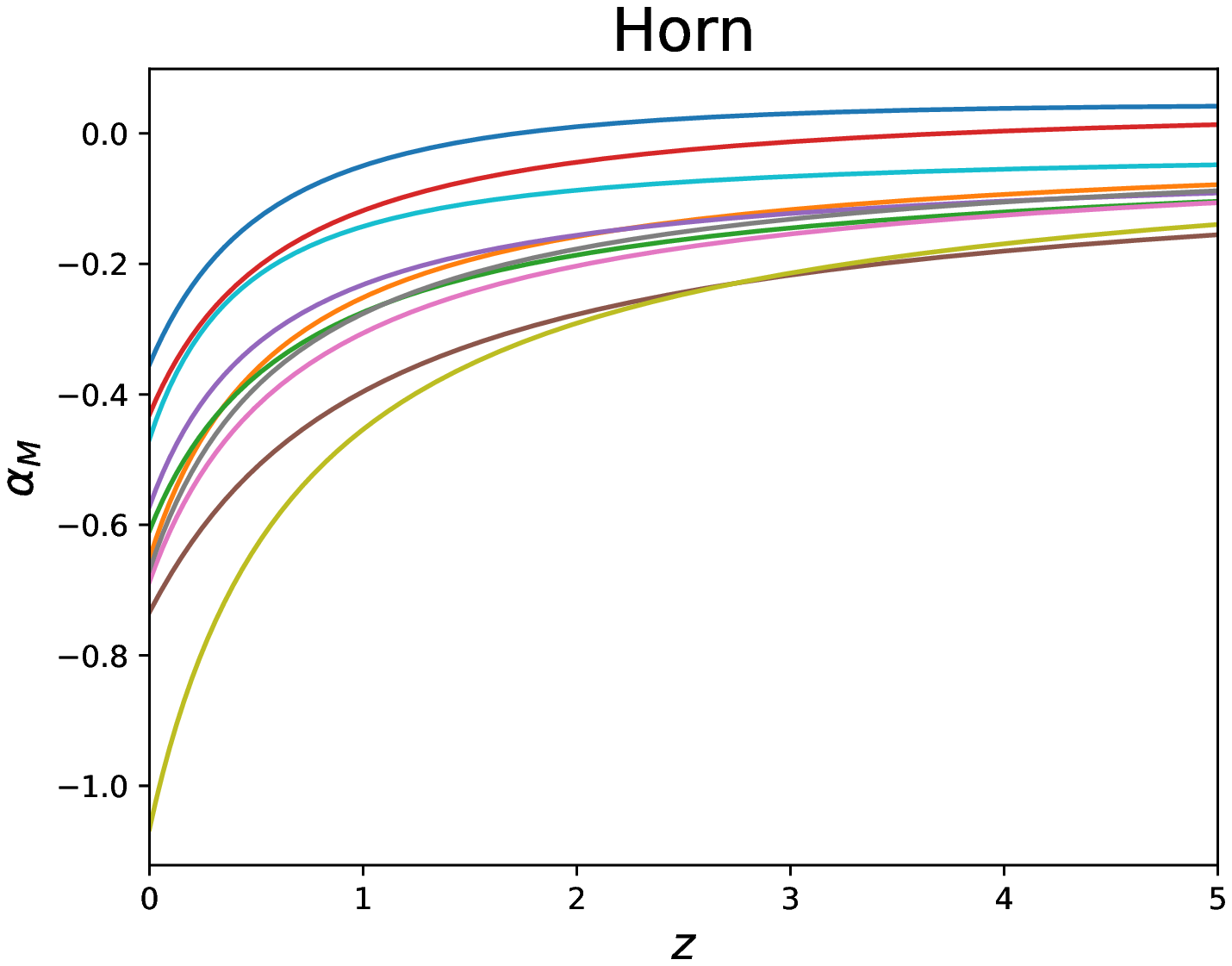}
    \includegraphics[width=8cm]{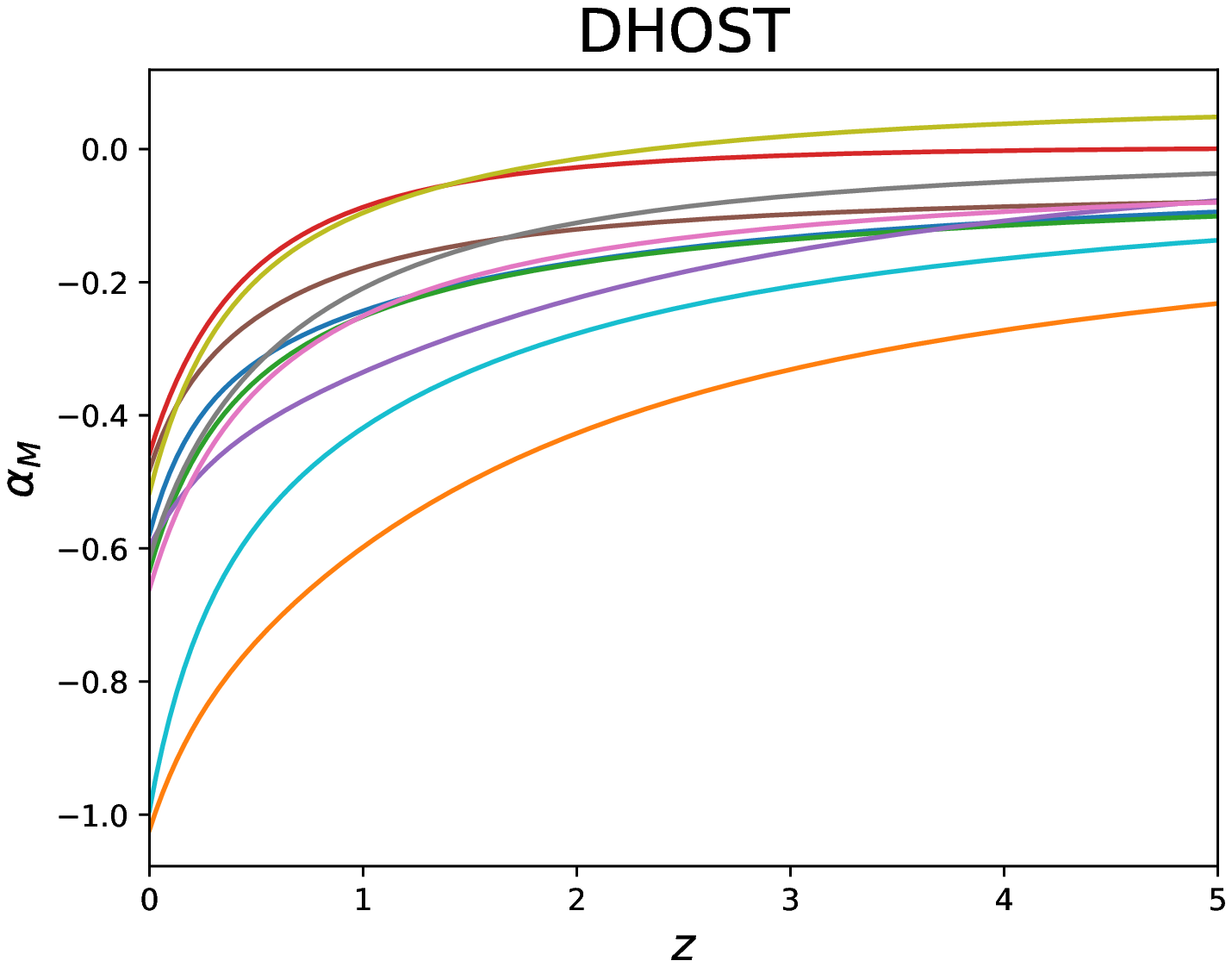} 
    \caption{The time evolution of $\alpha_M$ in Horndeski theory (left) and DHOST theory (right). 10 of $10^4$ generated models for each theory are arbitrarily selected. Note that the colors in the left and right figures do not correspond.}
   \label{fig:aM_timeev}
\end{center}
\end{figure}

\begin{figure}[h]
\begin{center} 
    \includegraphics[width=12cm]{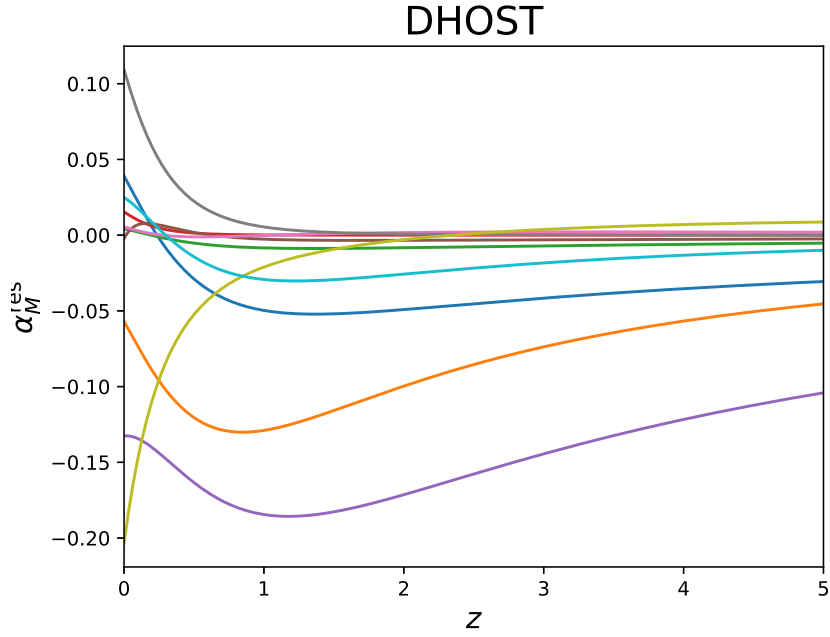} 
    \caption{The time evolution of $\alpha^{\rm res}_M$ in DHOST theory are plotted in the range of redshift $0\leq z\leq 5$. The color corresponds to the same models as in the right panel of Fig.~\ref{fig:aM_timeev}.} 
   \label{fig:aMres_timeev}
\end{center}
\end{figure}

Fig.~\ref{fig:aM_timeev} shows the time evolution of $\alpha_M$ as a function of the redshift for the Horndeski and DHOST theories. Although the $\alpha_M$ evolution apparently looks monotonic and shows similar shapes in both theories, it evolves differently in the low redshift. In Fig.~\ref{fig:aMres_timeev}, the measure of the deviation of the DHOST theory from the Horndeski theory, $\alpha^{\rm res}_M$, shows oscillations at $z \lesssim 2$. This low-$z$ behavior of $\alpha^{\rm res}_M$ eventually contributes to broadening the ranges of the $\alpha_M$ in the DHOST theory, especially at $z \lesssim 2$, as shown in the right panel in Fig.~\ref{fig:aM_timeev}. We find that $\alpha^{\rm res}_M$ stays negative in the models we constructed and converges to zero at higher redshifts, but quite slowly for some models.

\begin{figure}[h]
\begin{center} 
    \includegraphics[width=8cm]{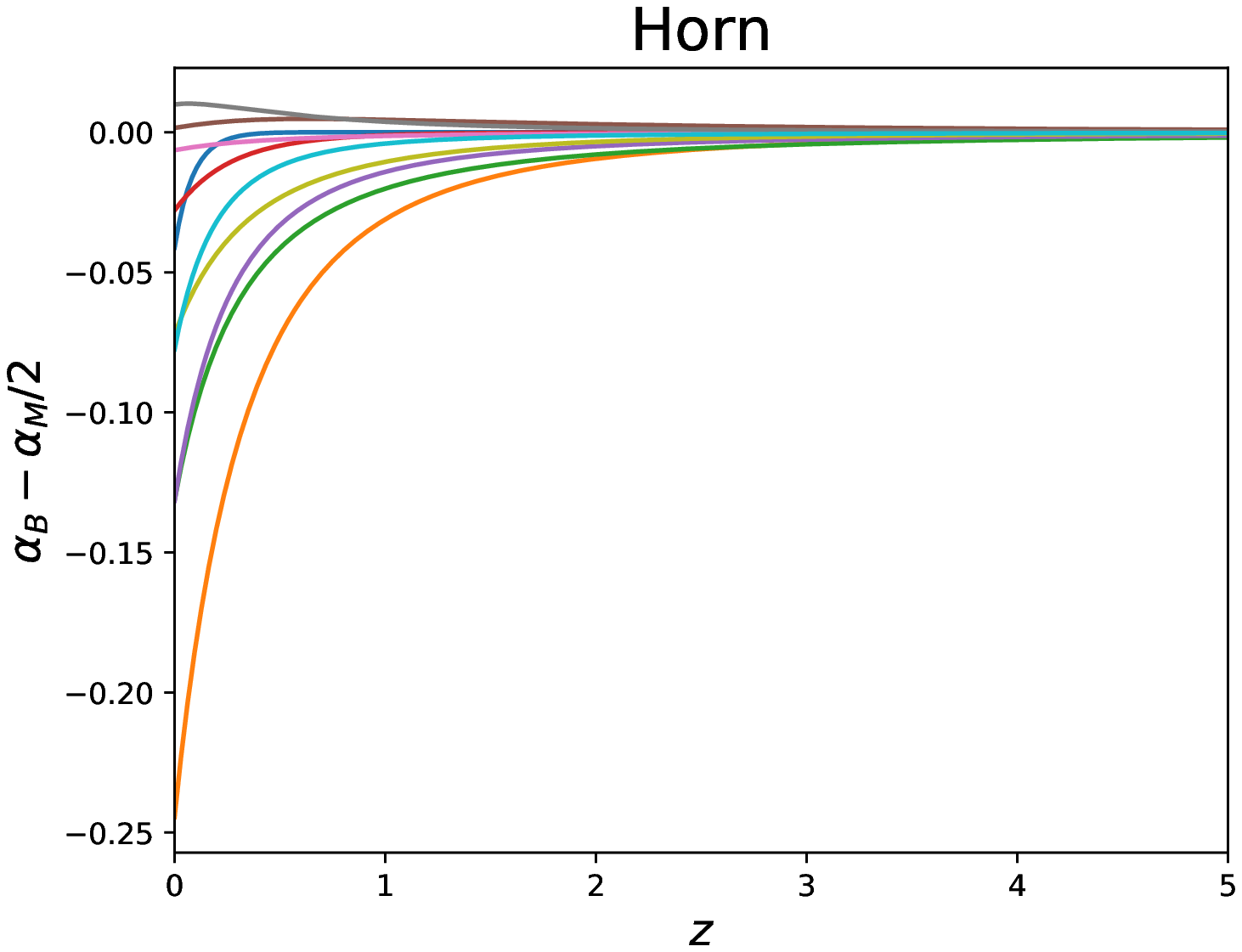} 
    \includegraphics[width=8cm]{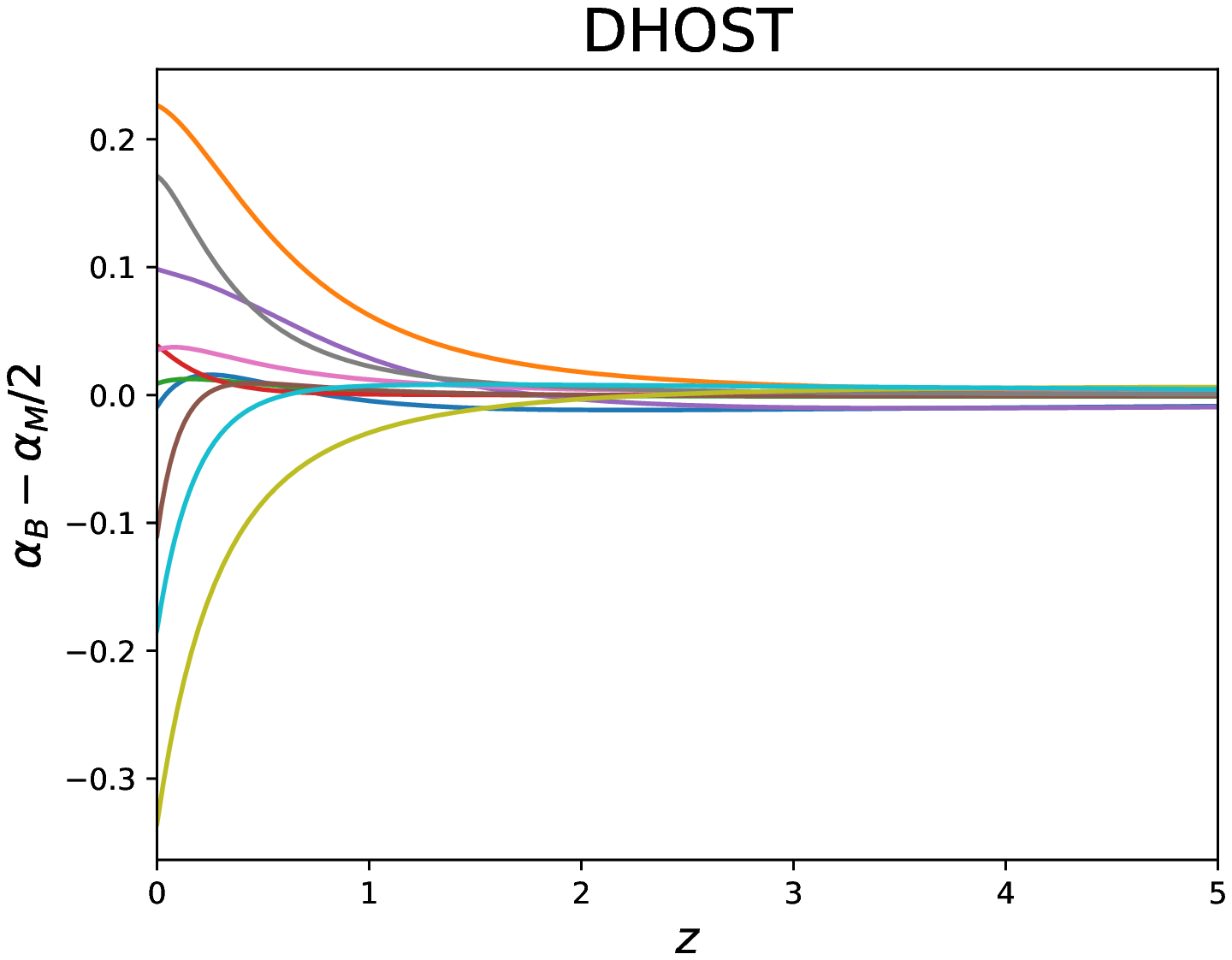} 
    \caption{The time evolution of $\alpha_B-\alpha_M/2$ in Horndeski theory (left) and DHOST theory (right). The color corresponds to the same models in each theory as in Fig.~\ref{fig:aM_timeev}.} 
   \label{fig:aB05aM_timeev}
\end{center}
\end{figure}

\begin{figure}[h]
\begin{center} 
    \includegraphics[width=12cm]{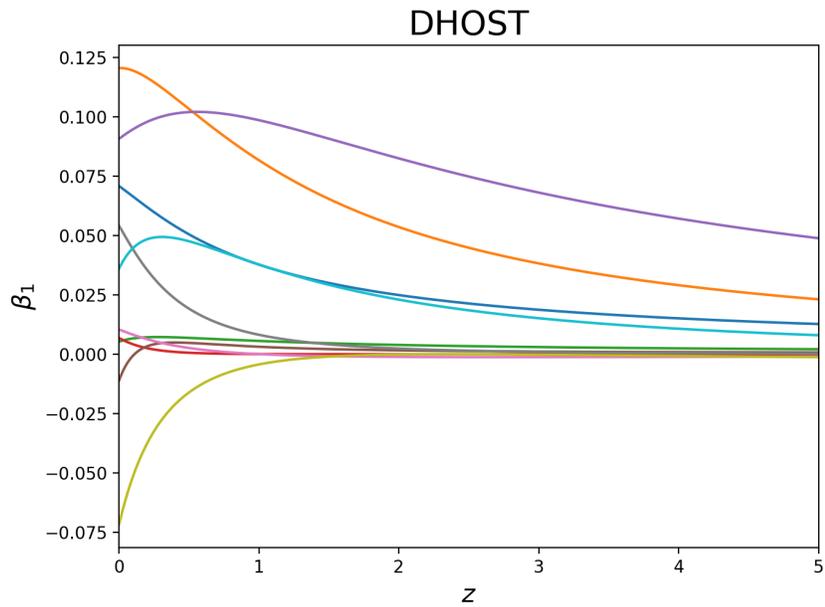} 
    \caption{The time evolution of $\beta_1$ in DHOST theory. 10 models of $10^4$ generated models are arbitrary selected. The color corresponds to the same models as in Fig.~\ref{fig:aM_timeev} and Fig.~\ref{fig:aB05aM_timeev}.} 
   \label{fig:b1_timeev}
\end{center}
\end{figure}

In Fig.~\ref{fig:aB05aM_timeev}, $\alpha_B-\alpha_M/2$ fluctuates at $z \lesssim 2$ and swiftly converges to zero at higher redshifts for both theories. $\alpha_B-\alpha_M/2$ at $z \lesssim 2$ in the DHOST theory oscillates more rapidly with larger amplitudes than the Horndeski counterpart, leading to the diversity of $\alpha_B-\alpha_M/2$ both in positive and negative directions. In Fig.~\ref{fig:b1_timeev}, $\beta_1$ also oscillates at low redshifts and slowly converge to zero at higher redshifts.

The broadened feature in $\alpha_M$ and the oscillatory ones in $\alpha_B-\alpha_M/2$ or $\beta_1$ are sourced via $X$ or $\dot{X}$ in Eqs.~\eqref{aMHorn-res}, \eqref{aB_res}, and \eqref{beta1}, which are generally expected in the DHOST theory while little in the Horndeski theory. 
Particularly, the oscillatory feature is generated through $X$ dependence of $F(\phi, X)$, which does not appear in the Horndeski theory after setting $c_g=c$.

In particular, $\alpha_B-\alpha_M/2$ significantly distinguishes the Horndeski theory and the DHOST theory in their redshift evolution.

\clearpage

\section{Discussion}\label{sec:dis}
In the following discussion, we comment on the impact of our results on the existing parameterization, and constraints on the DHOST theory.

\begin{itemize}
    \item The condition for evading graviton decay obtained in \cite{Creminelli:2018xsv} is $A_3=0$, i.e, $\alpha_H + 2\beta_1=0$. Indeed the A3eq0 theory is precisely the theory when we apply the constraint from the no graviton decay. 
    %\hl{In Fig.~\ref{fig:aM_aB_b1}, setting $A_3=0$ slightly shifts the distribution of the model parameters, but not significantly since the terms associated to $A_3$ in $\alpha_B$ and $\beta_1$, are small (of the order of ${\cal O}(|\hat{X}|^2)$ with $|\hat{X}| \ll 1$). This means that the constraint from the no graviton decay is insignificant to our claim on the features in Fig.~\ref{fig:aM_aB_b1}. Therefore, under the slow-rolling assumption $|\hat{X}| \ll 1$, we conclude that it is enough to consider the DHOST theory with the no graviton decay condition $A_3=0$ to capture the essential features of the DHOST theory at cosmological scales.} 
     However, the impact of graviton decay constraint is very insignificant, at least at cosmological scales, when the DE field is rolling slowly. Because, the terms associated to $A_3$ in $\alpha_B$ in Eq. \eqref{aB_res} and $\beta_1$ in Eq. \eqref{beta1}, are very small which is the order of ${\cal O}(|\hat{X}|^2)$ under the slow-rolling assumption, $|\hat{X}| \ll 1$. Indeed, Fig.~\ref{fig:aM_aB_b1} shows that the difference between A3eq0 (right) and DHOST (left) theories are very insignificant, i.e., $A_3=0$ leads to a slight shift towards the left in the distribution of the model parameters.
    
    %\textrightarrow 
    %\PKK{However, the impact of graviton decay constraint is very insignificant, at least at cosmological scales, when the DE field is rolling slowly. Because, the terms associated to $A_3$ in $\alpha_B$ in Eq. \eqref{aB_res} and $\beta_1$ in Eq. \eqref{beta1}, are very small which is the order of ${\cal O}(|\hat{X}|^2)$ under the slow-rolling assumption, $|\hat{X}| \ll 1$. Indeed, Fig.~\ref{fig:aM_aB_b1} shows that the difference between A3eq0 (right) and DHOST (left) theories are very insignificant, i.e., $A_3=0$ leads to a slight shift towards the left in the distribution of the model parameters.} 
    %\PK{a bit straight description. Please also clarify in the page 16.}\SA{Agreed. Please replace the statement as you meant.}
\end{itemize}

\begin{itemize}
    \item The remaining DHOST models after the constraint of the no graviton decay ($\alpha_H + 2\beta_1=0$) are principally characterized by $\alpha_M$, $\alpha_B-\alpha_M/2$ and $\beta_1$. Let us mention the current constraints on the present values of these parameters. $\alpha_M$ is currently bounded at small scales; $|\alpha_M|={\cal O}(10^{-2})$ \cite{Zhu:2018etc, Williams:2004qba,Hofmann:2018myc}, only when the screening mechanism are realized. $\alpha_M $ and $\alpha_B$ has been constrained in the Horndeski theory at cosmological scales \cite{Renk:2017rzu,Kreisch:2017uet,Noller:2018eht,SpurioMancini:2019rxy}, typically $|\alpha_M|, |\alpha_B| = {\cal O}(10^{-1})$, whereas has yet to be constrained in the DHOST theory. References \cite{Hirano:2019scf,Crisostomi:2019yfo} claimed that the measurement of the orbital decay rate of the Hulse - Taylor binary pulsars constrains up to $|\beta_1| = {\cal O}(10^{-3})$. Moreover, the simultaneous fitting of the X-ray and lensing profiles of galaxy clusters could reaches at $|\beta_1| = {\cal O}(10^{-1})$ as mentioned in \cite{Hirano:2019scf}. In our simulation, $|\beta_1| = {\cal O}(10^{-1})$ is allowed at lower redshifts as shown in Fig.~\ref{fig:b1_zev}. If we assume that $\beta_1$ at local scales could be extrapolated to cosmological scales, the Hulse-Taylor pulsar rules out almost all the extended Horndeski models in Fig.~\ref{fig:aM_aB_b1}. On the other hand, our models are still compatible with the constraint on $\beta_1$ from galaxy clusters.
    \end{itemize}

    \begin{itemize}
    \item Very recently, the paper \cite{Creminelli:2019kjy} claims that the instability of dark energy can be induced by the kinetic - braiding interaction in the system of a compact binary. The instability is evaded if the kinetic - brading term in the Lagrangian is dropped off. In the Horndeski theory, $Q(\phi,X) \sim 0$ is obtained, resulting in $\alpha_M - \alpha_B/2 \sim 0$ from Eqs.~\eqref{aM_Horn} and \eqref{aB_Horn}. In the DHOST theory, however, $\alpha_M-\alpha_B/2$ still deviates from zero due to the effects from $F(\phi,X)$, even after setting $Q(\phi,X) \sim 0$. This may indicate that the parameter $\alpha_B-\alpha_M/2$ is significant to probe the DHOST theory.
    \end{itemize}
    
    \begin{itemize}
    \item The redshift evolutions of $(\alpha_M,\alpha_B-\alpha_M/2,\beta_1)$ in the DHOST theory show the oscillatory features that are hardly realized in the Horndeski theory. The parametrization for the time evolution of $\alpha_M$ and $\alpha_B$ in cosmology is often assumed to be monotonic in literature, such as $\alpha_{M,B}(z) =\alpha_{M,B}(z=0)\times (1-\Omega_{\rm m}(z))$ in \cite{Langlois:2015cwa} or $\alpha_{M,B}(z)=\alpha_{M,B}(z=0) \times (1+z)^{-\beta}$ in \cite{Ade:2015rim,Kreisch:2017uet}. Such parametrizations may approximately work for the Horndeski theory as confirmed in this paper and \cite{Nishizawa:2019rra}, but is no longer valid in the DHOST theory because of the oscillations.
    \end{itemize}
    
After all, our predictions in the DHOST theory are still worth being tested by observations at cosmological scales. For observations, the cosmological perturbations need to be studied further in the DHOST theory, except for the linear growth of matter in the shift-symmetric case \cite{Hirano:2019nkz}. In addition, it is significant to take into account the oscillatory behavior of $\alpha_B-\alpha_M/2$, or $\beta_1$ to trace their redshift evolution and compare with observational data.

\section{Conclusion}\label{sec:conc}
We have numerically investigated the DHOST theory after GW170817, i.e., ${\rm DHOST}_{c_g^2=1}$ theory with the conventional matter at cosmological scales. We assumed the slow time evolution of the scalar field, $|\hat{X}| \ll 1$, particularly realizing the cosmic expansion of the late-time acceleration and the matter dominant epoch. We numerically computed the conventional EFT parameters, and found that the stable models that explain the cosmic acceleration within the ${\rm DHOST}_{c^2_g=1}$ theory framework have the following features:

\begin{itemize}

    \item The Planck mass run rate, $\alpha_M$, is almost identical in all subclasses of the DHOST$_{c^2_g=1}$ theory, which makes difficult to distinguish the DHOST$_{c^2_g=1}$ theory from the Horndeski theory. In general, $\alpha_M$ has a negative value, $\alpha_M \leq -0.1$, as found in the Horndeski theory in \cite{Nishizawa:2019rra}.

    \item The kinetic braiding parameter, $\alpha_B$, marginally distinguishes the theories if it is in the range of $\alpha_B \gtrsim 0$. Particularly, the Horndeski theory is clearly distinguished from the DHOST$_{c^2_g=1}$ theory.
    
    \item $\alpha_H$ and $\beta_1$ are correlated by $\alpha_H + 2\beta_1 = {\cal O}(|\hat{X}^2|)$, which is generically satisfied in the DHOST$_{c^2_g=1}$ theory. The values of $\alpha_H$ and $\beta_1$ in our computation are in the ranges, $-0.5 \lesssim \alpha_H \lesssim 0.2$ and $-0.1 \lesssim \beta_1  \lesssim 0.2$, respectively.  This consistently supports the relation $\alpha_H+2\beta_1 = {\cal O}(|\hat{X}^2|)$. 
    
    \item 
    The GLPV theory peculiarly predicts $\alpha_H ={\cal O}(|\hat{X}|^2)$, and deviates from the Horndeski theory in $\alpha_M$ and $\alpha_B$ at the order of ${\cal O}(|\hat{X}|^2)$. This is due to the condition of $\beta_1=0$. This makes the discrimination of these theories difficult.
    
\end{itemize}
In conclusion, we note that the correlations among $\alpha_M$, $\alpha_B$, and $\alpha_H$, and $\beta_1$ reduce the number of the characteristic parameters to three parameters. We propose that a parameter set of $(\alpha_M,\alpha_B-\alpha_M/2,\beta_1)$ is the principal set to discriminate the subdivision of the DHOST$_{c^2_g=1}$ theory. We find that the common parameters $\alpha_M$ and $\alpha_B-\alpha_M/2$ in the Horndeski and DHOST$_{c^2_g=1}$ theories can differ by the oscillatory features in their redshift evolutions. Our prediction on $(\alpha_M,\alpha_B-\alpha_M/2,\beta_1)$ can provide a broad opportunity to test the DHOST$_{c^2_g=1}$ theory for the cosmological surveys such as cosmic shear measurements \cite{Hildebrandt:2016iqg,Troxel:2017xyo,Hikage:2018qbn, Amendola:2012ys,Abate:2012za,Spergel:2013tha} and upcoming GW observations \cite{Nishizawa:2019rra,Belgacem:2019pkk}. We will address the constraints on $(\alpha_M,\alpha_B-\alpha_M/2,\beta_1)$ quantitatively from the different observations in the future work. 

%%%%%%%%%%%%%%%%%%%%%%%%%%%%%%%%%%%%%%%%%%
%%%%%%%%% Begin Acknowledgement %%%%%%%%%%
\begin{acknowledgments}
We would like to thank E.~Babichev, M.~Crisostomi,
T.~Hiramatsu, T.~Kobayashi, S.~Nojiri, I.~Sawicki, D.~Sorokin, G.~Tasinato, D.~Yamauchi, J.~Yokoyama, and M.~Zumalacarregui for useful discussions and comments on an early version of this manuscript and N.~Bartolo and S.~Matarrese for the warm hospitality during our stay at the University of Padova. P.~K sincerely expresses gratitude to the Kobayashi-Maskawa Institute for the Origin of Particles and the Universe (KMI), Nagoya University, for the visiting position and hospitality in Nagoya. S.~A. is supported by Research Fellow of the Japan Society for the Promotion of Science, No. 17J04978. P.~K acknowledges financial support from the Universit\`a di Roma - Progetti di Ricerca Medi 2017, prot. RM11715C81C4AD67, and the ``Fondazione Ing. Aldo Gini''. A.~N. is supported by JSPS KAKENHI Grant Nos. JP17H06358, JP18H04581, and JP19H01894. 
\end{acknowledgments}
%%%%%%%%%%%%%%%%%%%%%%%%%%%%%%%%%%%%%%%%%%
%%%%%%%%% Begin Appendix %%%%%%%%%%%%%%%%%
\newpage

\appendix

\section{FRW background equations}\label{app:frw:back}

The Friedman equations of the DHOST theory are
\begin{align}
&3M^2_*H^2 = \rho_m + \cal E\,, \label{F1}\\
&M^2_*(2\dot{H} + 3H^2) = -p_m - \cal P\,, \label{F2}
\end{align}
where the effective mass, $M^2_* = 2F$. $\rho_m$ and $p_m$ are the background energy density and pressure of all the matter components together, while $\cal E$ and $\cal P$ are the background energy density and pressure of the dark energy, which are defined as

\begin{eqnarray}\label{eom:energy}
{\cal E} &=&  -6\left(4F_X \dot{\phi}^2-\frac{3}{2}A_3\dot{\phi}^4\right) H^2 +3\left(-4{F_X} \dot \phi^2 +A_3 \dot{\phi}^4\right) \dot H  \nonumber\\ &&
 + 6 \left( - F_\phi \dot \phi + Q_X \dot \phi^3 + \frac{1}{2}A_{3\phi}\dot{\phi}^5+ \left( 2 {F_X} \dot \phi + \left( -\frac{1}{2} A_3 + 6 \frac{{F_X}^2}{F} \right) \dot \phi^3 - 3 A_3 \frac{{F_X}}{F} \dot \phi^5+ \frac{3}{8} \frac{A_3^2}{F} \dot \phi^7 \right) \ddot \phi \right) H  \nonumber\\ &&
 + 2 \left(6 \frac{{F_X}^2}{F} \dot \phi^3 - 3 A_3 \frac{{F_X}}{F} \dot \phi^5 + \frac{3}{8} \frac{A_3^2}{F} \dot \phi^7 \right) \dddot \phi \nonumber\\&&
+ \Bigg( 6 \frac{{F_X}^2}{F} \dot \phi^2 + \left( - 9 A_3 \frac{{F_X}}{F} + 12 \frac{{F_X}^3}{F^2} - 24 \frac{{F_X} {F_{XX}} }{F} \right) \dot\phi^4 \nonumber\\&&
+ 6 \left( \frac{5}{16}\frac{A_3^2}{F} + \frac{A_{3X} {F_X}}{F} - A_3 \frac{{F_X}^2}{F^2} + A_3 \frac{{F_{XX}} }{F}\right) \dot\phi^6 + \left( -\frac{3}{2} \frac{A_3 A_{3X}}{F} +\frac{3}{4} \frac{A_3^2 {F_X}}{F^2}\right) \dot\phi^8\Bigg)\ddot\phi^2 \nonumber\\&&
% %
\Bigg(\left( - 12 \frac{{F_X}^2 {F_\phi}}{F^2} + 24 \frac{{F_X} F_{\phi X}}{F} \right) \dot\phi^4 
\nonumber\\ &&+ 6 \left( -  \frac{A_{3\phi} {F_X}}{F} + A_3 \frac{{F_X} {F_\phi}}{F^2} - A_3 \frac{F_{\phi X}}{F}\right) \dot\phi^6 +\frac{3}{4} \left( 2 \frac{A_3 A_{3\phi}}{F} - \frac{A_3^2 {F_\phi}}{F^2}\right) \dot\phi^8\Bigg)\ddot\phi \nonumber\\&&
+ ( -2P_{X} + Q_\phi) \dot \phi^2 - P \,,\label{E} \\
{\cal P} &=& 4 \left( {F_\phi} \dot\phi - 2 {F_X} \dot \phi \ddot\phi\right) H +(- 4 {F_X} \dot \phi +A_3 \dot{\phi}^3)\dddot \phi\nonumber\\&&
 + \Bigg( - 4 {F_X} + \left(3A_3 - 6 \frac{F^{2}_X}{F} + 8 {F_{XX}} \right) \dot \phi^2 + \left(3 A_3 \frac{{F_X}}{F} - 2A_{3X}\right) \dot \phi^4 - \frac{3}{8} \frac{A_3^2}{F} \dot \phi^6 \Bigg) \ddot\phi^2 \nonumber\\&&
 + 2 \left( {F_\phi} - (4 F_{\phi X} + Q_X) \dot \phi^2+\frac{1}{2}A_{3\phi}\dot{\phi}^4\right) \ddot \phi + \left( 2 F_{\phi\phi} + Q_\phi\right) \dot \phi^2 + P \,. \label{P}
\end{eqnarray}
We indicate the appearence of $\dot{H}$ and $\dddot{\phi}$ in the above Friedmann equations. By using the spatial component of the Einstein equation, one can eliminate the $\dot{H}$ and $\dddot{\phi}$ in the temporal component if necessary, and rewrite the second order Friedmann equations as mentioned in \cite{Crisostomi:2019yfo,Crisostomi:2017pjs}. Here we are keeping $\dot{H}$ and $\dddot{\phi}$ in the equation without substitution because the higher derivatives would not give any trouble in our numerical computation.

%%%%%%%%%%%%%%%%%%%%%%
\section{Expansion of the scalar field}\label{app:taylor_expansion}
Here we derive the scalar field evolution given in Eq.~\eqref{e:phi:dim_less}. Since we assume the slowly varing scalar field, the scalar field $\phi$ can be expanded in the Taylor series in the conformal look back time, $\tau_{\rm LB}$, as
\begin{align}
\phi(\tau_{\rm LB}) = \sum^{N}_{n=0}{\frac{\phi^{(n)}(0)}{n !} {\tau_{\rm LB}}^n} \,,\label{e:phi}
\end{align}
where $\phi^{(n)} (0) \equiv{ \frac{d^n \phi}{d \tau^n_{\rm LB}}|}_{\tau_{\rm LB}=0}$, and $N$ is the truncation order of the Taylor series. We assume that $\phi$ varies slowly in the lower redshifts and almost constant in higher redshifts. Therefore, we truncate the seres in the third order, i.e., $N=3$. Hence we obtain the form in Eq.~\eqref{e:phi:dim_less}.
\begin{eqnarray}\label{phi_app}
\phi(\tau_{\rm LB}) = M_\phi \left \{ b_0+b_1 H_0 \tau_{\rm LB} + \frac{b_2}{2} (H_0 \tau_{\rm LB})^2 +\frac{b_3}{6}(H_0 \tau_{\rm LB})^3\right \}\,,
\end{eqnarray}
where $M_\phi$ is the mass scale of $\phi$ at present. Notice that the range of $b_i$ is still arbitrary. To determine $M_\phi$ and $b_i$, we expand the LB time given in Eq.~\eqref{phi_app} around $a=0$, i.e., the beginning of the Universe that we assume now. Then the expansion is not valid in the late-time Universe, $z<1$, but can be applied at least to the past of the Universe, $z>1$, including the era of the CMB recombination, $z\sim 10^3$. We use a prior knowledge that the matter dominates the Universe when $a \ll 1$. We are also assuming that the Hubble parameter is given by the $\Lambda {\rm CDM}$ model, which is approximated as $H \sim H_0\sqrt{\Omega_{\rm m0}a^{-3}(1+(a/a_t)^3)}$. By using it, we integrate and expand Eq.~ \eqref{tauLB} in Taylor series,
\begin{align}
\tau_{\rm LB}(a) =& \tau_{\rm LB}(0) - \frac{1}{H_0 a_t \sqrt{1-\Omega_{{\rm m0}}}} \times \left \{ 2{\left ( \frac{a}{a_t} \right )}^{1/2} + {\cal O}{\left (\left ( \frac{a}{a_t} \right )^{7/2} \right )}\right \} \,,\label{tau_app}
\end{align}
where $a_t \equiv (\Omega_{{\rm m0}}/(1-\Omega_{{\rm m0}}))^{1/3}$, i.e., the scale factor at matter and cosmological constant equality. Notice that the expansion in Eq.~\eqref{tau_app} only depends on $H_0$ and $\Omega_{\rm m0}$, regardless of the details of models we are interested in. Hereafter, we set the Hubble constant to $H_0 = 67.8\,{\rm km}\,{\rm s}^{-1}\,{\rm Mpc}^{-1}$, and the total matter density parameter including the cold dark matter and baryons to $\Omega_{m0}=0.3080$, as obtained by the Planck observation 2015 \cite{Ade:2015xua}. After inserting $\tau_{\rm LB}$ given in Eq.~\eqref{tau_app} into Eq.~\eqref{phi}, the $\phi$ is expressed in the scale factor $a$. $\phi(a)$ is then given as 
\begin{align}
\phi(a)/\tilde{M}_\phi = c_0 + \sum^N_{i=1}{ c_i (1-a^{i/2})}\,.\label{e:phi:dim_less_app}
\end{align}

Then we normalize $\phi$ at the limit of the early Universe so that
\begin{equation}\label{phi_ini}
    \phi(\tau_{\rm LB}(a=0))=\tilde{M}_\phi\,. 
\end{equation}

From Eq.~(\ref{phi_app}), we can calculate an asymptotic time evolution of $\phi$ as follows. From Eq.~(\ref{phi_app}), the cosmic time derivative of $\phi$, $\dot{\phi}$, is derived as $\dot{\phi} = -a^{-1}d\phi/d\tau_{\rm LB} = a^{-1}H_0 (b_1+b_2H_0\tau_{\rm LB})$. For the matter dominant epoch, i.e., $H^2 \propto a^{-3}$, we obtain $\dot{\phi}/H \propto H^{-1/3}$. This means that the time variation of $\phi$ becomes smaller in the past of the Universe. 

%%%%%%%%%%%%%%%%%%%%%%
\section{The derivation of the EFT parameters and stability conditions}\label{app:EFT}

Here we introduce the EFT parameters, and the stability conditions in the class Ia DHOST theory \footnote{The main arguments should be applicable to the more general class of the DHOST theory such that $A_1 = -A_2 \neq 0$.}

\subsection{EFT description of the Class Ia DHOST theory}\label{sapp:EFT_comp}
The metric in the ADM form reads,
\begin{align}
ds^2 = -N^2dt^2 + g_{ij}(dx^i+N^idt) (dx^j+N^jdt)\,, \label{ADMmetric}
\end{align}
where $N$ is the lapse and $N_i$ the shift vector. We define a time-like vector orthogonal to the foliation, $n^\mu$, as $n^\mu \partial_\mu = (1/N, -N^i/N)$. 
 We take the time-like vector $n^\mu$ proportional to the gradient of $\phi$, 
\begin{align}
\nabla_\mu \phi = -An^\mu\,, \label{phi_cloclk}
\end{align}
where $A \equiv n^\mu \nabla_\mu \phi$. $V$ is defined as the time derivative of $A$ as
\begin{align}
V \equiv n^{\mu}\nabla_\mu A\,. \label{V}
\end{align} 
The total action of \eqref{ac:total} and \eqref{ac:dhost:gw} is given by
\begin{align}
&S = \int{dtd^3xN{\sqrt g}{\cal L}}\,,\label{S0} \\
&{\cal L} = P + Q_2AK + F (R+K_{ij}K^{ij} - K^2) -2F_\phi AK + \left[ (A_3+A_4)X + A_5 X^2\right ] V^2 \cr 
& \qquad \qquad + (4F_X + A_3 X)AKV + (-4F_X  + A_4 X)\partial_i A \partial^i A \,, \label{L_DHOST}
\end{align}
where $g = {\rm det}{[g_{ij}]}$ and $Q_2(\phi,X)$ satisfies $Q = Q_1+2XQ_{1X}$ with $Q_1 \equiv \frac{1}{2}\int{dX(-X)^{3/2}Q_2(\phi,X)}$. We choose the unitary gauge $\phi=t$, which leads to $\nabla_i \phi = 0$, and expand around the FLRW metric, i.e., $ds^2 = -dt^2 +a^2\delta_{ij}dx^idx^j$. By following the notation in \cite{Langlois:2017mxy} the quadratic action for the EFT description is given as
\begin{align}
&S^{(2)}_{\rm EFT} = \int{dtd^3x a^3\frac{M^2}{2} \Biggl \{ \delta K_{ij} \delta K^{ij} - \delta K^2 + \left (\frac{\delta \sqrt{h}}{a^3}R+\delta_2 R \right )+  (1+\alpha_H)R\delta N}  \cr
&+ H^2\alpha_K \delta N^2 + 4H\alpha_B \delta N \delta K + 4\beta_1 \delta K \delta V  + \beta_2 \delta V^2 + \beta_3 v_iv^i\Biggr \}\,, \label{S2}
\end{align}
where $\delta V$ and $a_i$ are given as
\begin{align}
    &\delta V \equiv (\delta \dot{N} - N^i\partial_i N)/N\,,\label{deltaV}\\
    &v_i \equiv \partial_iN/N\,.\label{ai}
\end{align}
Since the second term of $\delta V$ in Eq.~\eqref{deltaV} is at the second order of the perturbations, the relation $\delta V = \delta \dot{N}$ is enough for computing the EFT parameters. As a consequence of the degeneracy conditions, $\beta_2$ and $\beta_3$ must satisfy the following conditions,
\begin{align}
\beta_2 = -6\beta^2_1\,,\ \beta_3 = -2\beta_1[2(1+\alpha_H)+\beta_1]\,.
\end{align}
Here we derive $\alpha_B$ and $\alpha_K$ in the following way. In the unitary gauge, $A=\dot{\phi}/N$ and $V=\ddot{\phi}/N^2$ at the background, both of which contains lapse function. Provided $N = \bar{N}+\delta N$, we obtain the perturbed $V$ as
\begin{align}
    V = \frac{\ddot{\phi}}{\bar{N}^2}\left(1-2\frac{\delta N}{\bar{N}} \right) - \frac{\dot{\phi}}{\bar{N}^2}\delta \dot{N}\,. \label{Vexp}
\end{align}
Note that the second term in the first bracket in Eq.~\eqref{Vexp} contributes to the perturbation of the Lagrangian associated with the lapse function, consequently changing $\alpha_{K,B}$. Importantly, the last term in Eq.~\eqref{Vexp} does not only appear with $\beta_{1,2,3}$, but also with $\alpha_K$ by the cross multiplication of the second term in the bracket and the last terms in $V^2$. We discuss this more specifically in the next paragraph. Hereafter we set $\bar{N}=1$. 

To obtain the explicit forms of $\alpha_{K,B}$ from the Lagrangian in Eq. \eqref{L_DHOST}, we apply the same computational strategy given in \cite{Gleyzes:2014dya}. According the expansion shown in Eq.~(\ref{S2}), $\alpha_B$ is formally given as
\begin{align}
\alpha_B = \frac{2H{\cal L}_{SN} + {\cal L}_{KN}}{4H{\cal L}_{\cal S}}\,, \label{alphaB_gen}
\end{align}
where ${\cal L}_a \equiv = \partial {\cal L}/\partial a$ and ${\cal S} \equiv K_{ij}K^{ij}$. The straightforward computation of Eq.~(\ref{alphaB_gen}) with the choice $N=1$ gives an explicit result,  Eqs.~(\ref{aB_Horn}) to (\ref{aB_res}). 
$\alpha_K$ on the contrary is more subtle to be computed.
During the perturbation in terms of $\delta_N$ from Eq.~(\ref{L_DHOST}) to Eq.~(\ref{S2}), the term $\delta N \delta \dot{N}$ appears from the term icluding $V^2$ and $AV$. The partial integral on this term, provides the additional terms in $\alpha_K$. The contribution from $\delta N \delta \dot{N}$ appears in the second term of the following equation,
\begin{align}
\alpha_K = \frac{2{\cal L}_N + {\cal L}_{NN}}{2H^2L_{\cal S}} - \frac{\dot{B} + 3HB}{H^2 {\cal L}_{\cal S}}\,, \label{alphaK_gen}
\end{align}
where $B$ is given as
\begin{align}
&B \equiv -2\dot{\phi}\ddot{\phi}X \left \{ 3(A_3+A_4)+4A_5X + (A_{3X}+A_{4X})X + A_{5X}X^2 \right \} \cr
&\qquad \qquad \qquad- 3HX \left \{ (4F_{XX}+A_{3X}X + A_3)X+\frac{3}{2}(4F_X+A_3 X)\right \}\,. \label{B_alphaK}
\end{align}
Notice that the GLPV theory, i .e., $4F_X+A_3 X=0$, $A_3+A_4=0$ and $A_5=0$, leads to $B=0$.
In the conformal frame where we are working, the form of $\alpha_K$ and $\alpha_B$ become complicated because $\delta N\delta \dot{N}$ exists by choice of the conformal frame such that the scale factor obeys the Friedmann equations in Eqs.~(\ref{F1}) and (\ref{F2}).

By computing the first term in Eq.~(\ref{alphaK_gen}), we obtain,
\begin{align}
&\alpha^{\rm Horn}_K =  \frac{1}{H^2F}\left \{X(P_X+2XP_{XX}-Q_\phi -2XQ_{\phi X})-6\dot{\phi}HX(Q_{X}+XQ_{XX}) \right \}\,, \label{aK_horn}\\  &\alpha^{\rm res}_K = -\frac{12(XF_X+4X^2F_{XX})}{F} -\frac{12\dot{\phi}X(3F_{\phi X}+2XF_{\phi XX})}{HF} \cr
& \qquad \quad -2V^2\left( 2\tilde{\beta_2} + 5X\tilde{\beta_2}_X +2X^2\tilde{\beta_2}_{XX}\right) + 6HAV\left(3\tilde{\beta_1}-3X\tilde{\beta_1}_X +2X^2\tilde{\beta_1}_{XX}\right) \cr
&\qquad \quad- \frac{\dot{B} + 3HB}{2H^2F}
\label{aK_res}\,,\nonumber
\\ 
\end{align}
where we define $\tilde{\beta_1} \equiv 4F\beta_1/X$ and $\tilde{\beta_2} \equiv F\beta_2/X$.

\subsection{Stability conditions in the absence of matters}\label{sapp:stab_wom}
Here we derive the stability conditions for the scalar and tensor perturbation. We start with the metric perturbation in the scalar sector. The metric is given as
\begin{align}
g_{00} = -(1 + \delta N)^2,\  g_{0i} = g_{i0} = a^2\partial_i \chi,\  g_{ij} = a^2(1+2\zeta)\delta_{ij}\,.\label{metric_perturb}
\end{align}
In the absence of matter, the quadratic action is
\begin{align}
&S^{(2)} = \int{dtd^3xa^3\frac{M^2_*}{2}\Biggl \{-6\dot{\zeta}^2 + 12 \beta_1 \dot{\zeta} \delta \dot{N} + \beta_2 \delta \dot{N}^2 +12H\left [ (1+\alpha_B)\dot{\zeta} - \beta_1 \delta \dot{N}\right] \delta N } \cr
& \qquad \qquad H^2(\alpha_K - 6 -12\alpha_B)\delta N^2 + 4\left [ \dot{\zeta}-\beta_1 \delta \dot{N} -H(1+\alpha_B)\delta N\right ]\partial^2 \chi \cr
& \qquad \qquad  \qquad \qquad \frac{1}{a^2} \left [ 2(1+\alpha_T)(\partial_i \zeta)^2 + 4(1+\alpha_H)\partial_i \zeta \partial_i \delta N + \beta_3 (\partial_i\delta N)^2\right ]\Biggr \}
\,. \label{S2_zeta}
\end{align}

The scalar perturbation is diagonalized with the quantity
\begin{align}
\tilde{\zeta} \equiv \zeta -\beta_1\delta N\,, \label{tildezeta}
\end{align}
and Eq.~\eqref{S2_zeta} becomes
\begin{align}
S_{\tilde{\zeta}} = \int{dtd^3xa^3\frac{M^2_*}{2}\left[ A_{\tilde{\zeta}}\dot{\tilde{\zeta}}^2 + B_{\tilde{\zeta}} \frac{(\partial_i {\tilde{\zeta}})^2}{a^2}\right]}\,, \label{S2_tildezeta}
\end{align}
where $\psi$ is the curvature perturbation in the spatial metric. Notice that $\tilde{\zeta}$ is not gauge invariant quantity because of existing $\delta N$. Basic quantities that appear in the action in Eq.~(\ref{S2_tildezeta}) are the coefficient on the kinetic terms and on the gradient term, $A_{\tilde{\zeta}}$ and $B_{\tilde{\zeta}}$, respectively. In the class Ia  DHOST theory $A_{\tilde{\zeta}}$ and $B_{\tilde{\zeta}}$ are given as,

\begin{align}
&A_{\tilde{\zeta}} = \frac{1}{(1+\alpha_B-\dot{\beta}_1/H)^2}\left [ \alpha_K + 6\alpha^2_B-\frac{6}{a^3H^2M^2_*}\frac{d}{dt}(a^3HM^2_*\alpha_B\beta_1)\right ] \,, \label{Azeta_wo_m}\\
&B_{\tilde{\zeta}} = 2-\frac{2}{aM^2_*}\frac{d}{dt}\left[ \frac{aM^2_*(1+\alpha_H+\beta_1)}{H(1+\alpha_B)-\dot{\beta}_1}\right]\,, \label{Bzeta_wo_m}\\
&C_{\tilde{\zeta}} = 0\,, \label{Czeta_wo_m}\\
&\alpha =  \alpha_K + 6\alpha^2_B-\frac{6}{a^3H^2M^2_*}\frac{d}{dt}(a^3HM^2_*\alpha_B\beta_1)\,. \label{alpha}
\end{align}
To have the positive definite linear Hamiltonian, the scalar perturbation must obey
\begin{align}
    A_{\tilde{\zeta}} >0\,,\ B_{\tilde{\zeta}}<0\,.\label{stab_no_matter}
\end{align}

The spatial part of the metric is relevant for the tensor sector,
\begin{align}
g_{ij} = a^2(\delta_{ij}+h_{ij})\,.\label{metric_tensor}
\end{align}
The quadratic action for the tensor sector corresponding to the action Eq.~\eqref{S0} is

\begin{align}
    S^{(2)}_h = \int{dtd^3xa^3\frac{M^2_*}{2}\left[\dot{h}^2_{ij} - (\partial_k h_{ij})^2 \right]}\,. \label{S2_tensor}
\end{align}
The condition for avoiding the ghost instability for the tensor perturbation is 
\begin{align}
    M^2_*>0\,. \label{stab_M2}
\end{align}

Accounting the matter is inevitable for explaining the late time Universe. In other words, it is necessary to derive the stability conditions by including a matter other than the condition in Eq.~(\ref{stab_no_matter}).

\subsection{Gradient instability in the presence of conventional matter}
We assume a matter component we look into is described by a barotropic perfect fluid, i.e., $p_m = p_m(\rho_m)$. The behavior of a barotropic perfect fluid is well mimiced by a massless scalar field minimally coupled to gravity \cite{Boubekeur:2008kn}. Although the detailed physical property of a massless scalar field does not always exactly the same as that of a perfect fluid at certain situations \cite{DeFelice:2009bx}. In our paper, we consider a massless scalar field as a conventional matter by assuming in matching situations discussed in \cite{Boubekeur:2008kn}. 

According to Gleyzes et. al. \cite{Gleyzes:2014qga}, the stablity conditions of the GLPV theory are different from the Horndeski theory by nonzero $\alpha_H$. On top of that, the stability conditions of the DHOST theory are also distinguishable from the GLPV theory. Here we argue the stability condition of the DHOST theory in the presence of the convensional matter described by a scalar field, $\sigma$, minimally couples to gravity as 
\begin{align}
S_m = \int{d^3xdt}N\sqrt{h}P(\sigma, Y)\ , Y \equiv g^{\mu\nu}\partial_\mu \sigma \partial_\nu \sigma = -\frac{(\dot{\sigma}-N^{i}\partial_i \sigma)^2}{N^2} + h^{ij}\partial_i \sigma \partial_j \sigma\,.\label{action_sigma}
\end{align}
Notice that the inhomogeneity of $\sigma$ exists in the unitary gauge. Then {the matter field perturbed as} $\sigma = \sigma_0 + \delta \sigma$, which leads to the quadratic order perturned matter action for Eq.~\eqref{action_sigma}
\begin{align}
&S^{(2)}_m = \int{d^3x dt a^3\Biggl \{ \frac{\delta \sqrt{h}}{a^3}\delta N P + \left ( \frac{\delta \sqrt{h}}{a^3}+\delta N\right )(P_Y \delta_1 Y + P_\sigma \delta \sigma)} \cr
 & \qquad \qquad \qquad \qquad +P_Y \delta_2 Y + \frac{P_{YY}}{2}\delta_1 Y^2 + P_{Y\sigma}\delta_1 Y \delta \sigma + \frac{P_{\sigma \sigma} \delta \sigma^2}{2}\Biggr \}\,, \label{matter_2nd}
\end{align}
with 
\begin{align}
&\frac{\delta \sqrt{h}}{a^3} = 3\zeta\,, \label{sqrth}\\
&\delta_1Y = 2\dot{\sigma}^2_0\delta N - 2\dot{\sigma}_0\delta \dot{\sigma} \,, \label{dY}\\
&\delta_2 Y =  -3\dot{\sigma}^2_0\delta N^2-\delta \dot{\sigma}^2 + 4\dot{\sigma}_0 \delta \dot{\sigma} \delta N + 2\dot{\sigma}_0 \partial_i B \delta \partial^i \sigma + h^{ij}\partial_i \sigma \partial_j \sigma\,. \label{d2Y}
\end{align}
In the presence of the matter, the momentum constraint reads 
\begin{align}
\delta N = \frac{1}{H(1+\alpha_B)-\dot{\beta}_1}\left ( \dot{\tilde{\zeta}} + \frac{\rho_m + p_m}{2M^2}\frac{\delta \sigma}{\dot{\sigma}_0} \right )\,. \label{momentum_w_Y}
\end{align}
Then we introduce the quantity $Q_\sigma \equiv \delta \sigma  - (\dot{\sigma}_0/H)\tilde{\zeta}$. Note that $Q_\sigma$ is not a gauge invariant variable if $\beta_1 \neq 0$, eliminating $\delta N$.
Inserting $\delta N$ into Eq.~\ref{momentum_w_Y} and rewriting in terms of $\tilde{\zeta}$ and $Q_\sigma$, the whole quadratic action of gravity and matter reads
\begin{align}
S^{(2)} = \int{dt d^3x a^3 \left ( {\tilde{\cal L}}_{\tilde{\zeta}} + {\tilde{\cal L}}_{Q_\sigma} + \tilde{{\cal L}}_{\tilde{\zeta}Q_\sigma} + ({\rm non\ derivative\ terms})\right)}\,,\label{S2-whole}
\end{align}
with
\begin{align}
& {\tilde{\cal L}}_{\tilde{\zeta}} = \frac{M^2_*}{2}\left \{\tilde{A}_{\tilde{\zeta}}\dot{\tilde{\zeta}}^2 + \tilde{B}_{\tilde{\zeta}} \frac{(\partial_i \tilde{\zeta})^2}{a^2} \right \}\,, \label{tL_tzeta} \\
& {\tilde{\cal L}}_{Q_\sigma} =  -\frac{P_Y}{c^2_m} \left ( \dot{Q}^2_\sigma  - c^2_m \frac{(\partial_i Q_\sigma)^2}{a^2}\right )\,, \label{tL_Qsigma}\\
& \tilde{{\cal L}}_{\tilde{\zeta}Q_\sigma} = -\frac{2\dot{\sigma_0}P_Y}{c^2_m(H(1+\alpha_B)-\dot{\beta}_1)}\left ( (\alpha_B -\dot{\beta}_1/H)\dot{\tilde{\zeta}}\dot{Q}_\sigma - c^2_m(\alpha_B-\dot{\beta}_1/H-\alpha_H-\beta_1)\frac{\partial_i \tilde{\zeta} \partial_i Q_\sigma}{a^2}\right )
\end{align}

\begin{align}
&\tilde{A}_{\tilde{\zeta}} = A_{\tilde{\zeta}} + \frac{(\rho_m + p_m)}{H^2 M^2_* c^2_m}\left ( \frac{H\alpha_B - \dot{\beta}_1}{H(1+\alpha_B)-\dot{\beta}_1}\right )^2\,,\label{Azeta_w_Y}\\
&\tilde{B}_{\tilde{\zeta}} = B_{\tilde{\zeta}} -\frac{\rho_m + p_m}{M^2_*H^2}\left (1 - \frac{2(1+\alpha_H+\beta_1)}{1+\alpha_B-\dot{\beta}_1/H}\right ) \label{Bzeta_w_Y}\\
&\tilde{C}_{\tilde{\zeta}} = C_{\tilde{\zeta}} =0\,.\label{Czeta_w_Y}
\end{align}
 Here $\rho_m + p_m = -2\dot{\sigma}^2_0 P_Y $, and the sound speed of the matter is $c^2_m \equiv P_Y/(P_Y - 2\dot{\sigma}^2_0 P_{YY})$. 
We rewrite the quadratic action in Eq.~(\ref{S2-whole}) as
 \begin{eqnarray}
 S^{(2)}=\int{dt d^3x a^3\frac{M^2_*}{2}\left({\dot{\mathbf{x}}}^T{\cal K}{\dot{\mathbf{x}}}+\frac{{\partial_i {\mathbf{x}}^T}{\cal G}{\partial_i {\mathbf{x}}}}{a^2} \right)}\,,
 \end{eqnarray}
 where $\mathbf{x} \equiv (\tilde{\zeta},Q_\sigma)$, and
 
 \begin{eqnarray}
  {\cal K} = \left (
 \begin{array}{cc}
 \tilde{A}_{\tilde{\zeta}} & A(\alpha_B-\dot{\beta}_1/H)\\
 A(\alpha_B-\dot{\beta}_1/H) & -2P_Y/M^2_*c^2_m
 \end{array}
 \right )
 \,, \label{calK}
 \end{eqnarray}
 
 \begin{eqnarray}
 {\cal G} = \left (
 \begin{array}{cc}
  \tilde{B}_{\tilde{\zeta}} &  -Ac^2_m(\alpha_B-\dot{\beta}_1/H - \alpha_H - \beta_1) \\
 -Ac^2_m(\alpha_B-\dot{\beta}_1/H - \alpha_H - \beta_1)  & 2P_Y/M^2_*
 \end{array}
 \right )
 \,, \label{calG}
 \end{eqnarray}
with
 \begin{eqnarray}
 A = \frac{-2\dot{\sigma}_0P_Y}{H M^2_*c^2_m (1+\alpha_B-\dot{\beta}_1/H)}\,.
 \end{eqnarray}
 To avoid the ghost and gradient instabilities of a cosmological solution, the eigenvalues of ${\cal K}$ must be positive, and the eigenvalues of ${\cal G}$ must be negative. Since ${\cal K}$ and ${\cal G}$ are a symmetric matrix, the necessarry and sufficient conditions of the stability is
\begin{eqnarray}
{\rm Tr}({\cal K}) > 0 \ {\rm and}\  {\rm det}({\cal K})>0\,,\label{calK-posi} \\
{\rm Tr}({\cal G}) < 0 \ {\rm and}\  {\rm det}({\cal G})>0\,.\label{calg_nega}
\end{eqnarray}
Eqs.~(\ref{calK-posi}) and (\ref{calg_nega}) with the null energy condition of the matter, i.e., $P_Y<0$ leaves the condition
\begin{align}
    A_{\tilde\zeta} >0,\  B_{\tilde{\zeta}}+\frac{\rho_m+p_m}{M^2_*H^2}\left( \frac{1+\alpha_H+\beta_1}{1+\alpha_B-\dot{\beta}_1/H}\right)^2 < 0\,.\label{stab_wm1}
\end{align}
Note that one can recover the stability condition in the absence of matter in Eq.~\eqref{stab_no_matter} from the above equation , i.e, decoupling limit of the matter from gravity.

The stability conditions of the DHOST theory in the presence of matter have been derived in ref. \cite{Crisostomi:2019yfo}. However, their conditions is slightly different than us in Eq.~(\ref{stab_wm1}). The conditions in Eq.~(\ref{stab_wm1}) is continuously applicable toward the super horizon region, described by the initial conditions on $\tilde{\zeta}$ and $Q_\sigma$. In fact, $\tilde{\zeta} $ and $Q_\sigma$ recovers their gauge invariance in the case of the GLPV, i.e., $\beta_1=0$. In fact, the conditions in the paper \cite{Crisostomi:2019yfo} and Eq.~\eqref{stab_wm1} leads the same expression in the limit of $\beta_1=0$. However, we admit that the variation of the stability conditions is crucial for cosmology. 

\subsection{Basis dependency of the linear stability conditions}
\label{sapp:linear_basis}

\begin{figure}[h]
\begin{center} 
    \includegraphics[width=10.0cm]{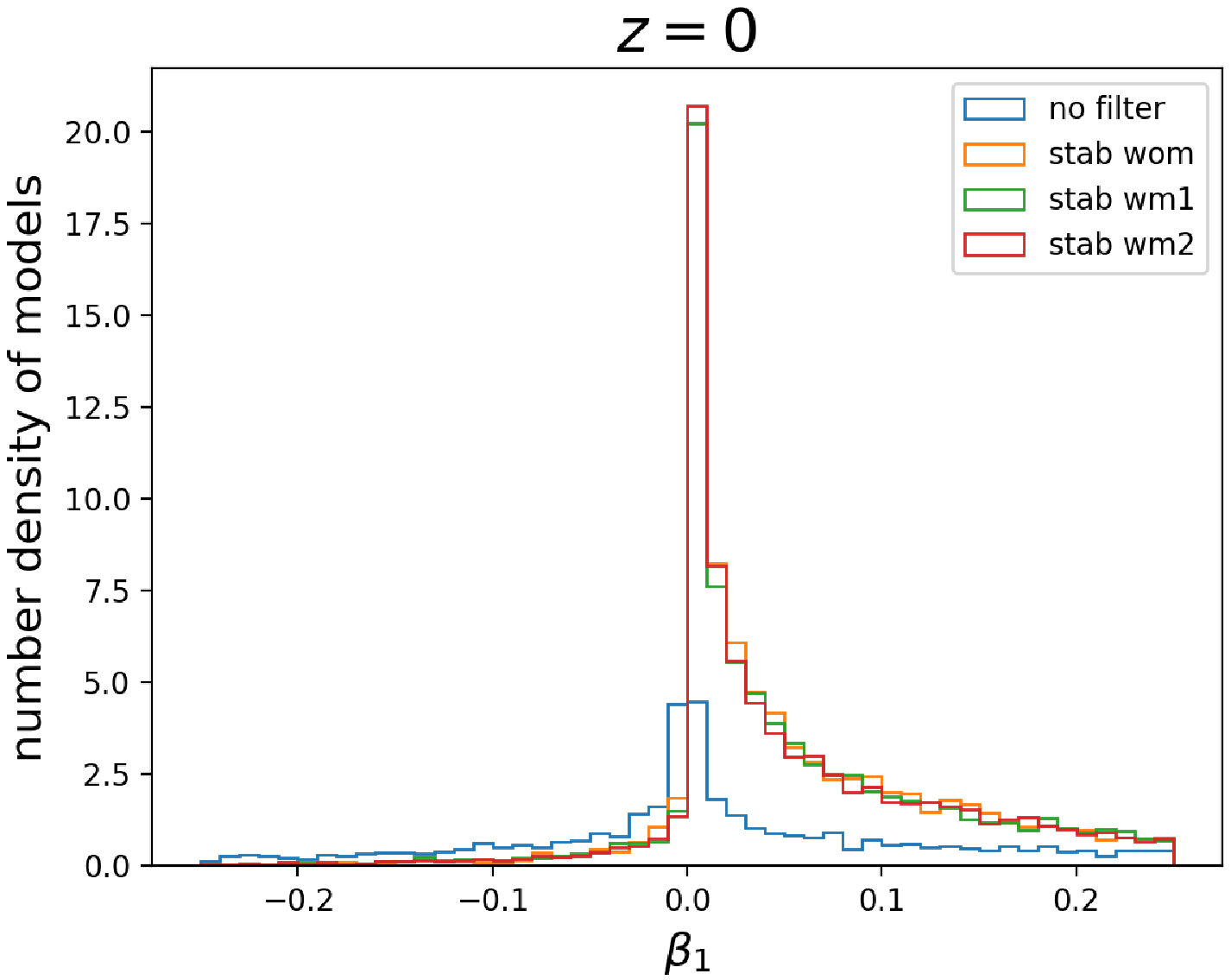} 
    \includegraphics[width=10.0cm]{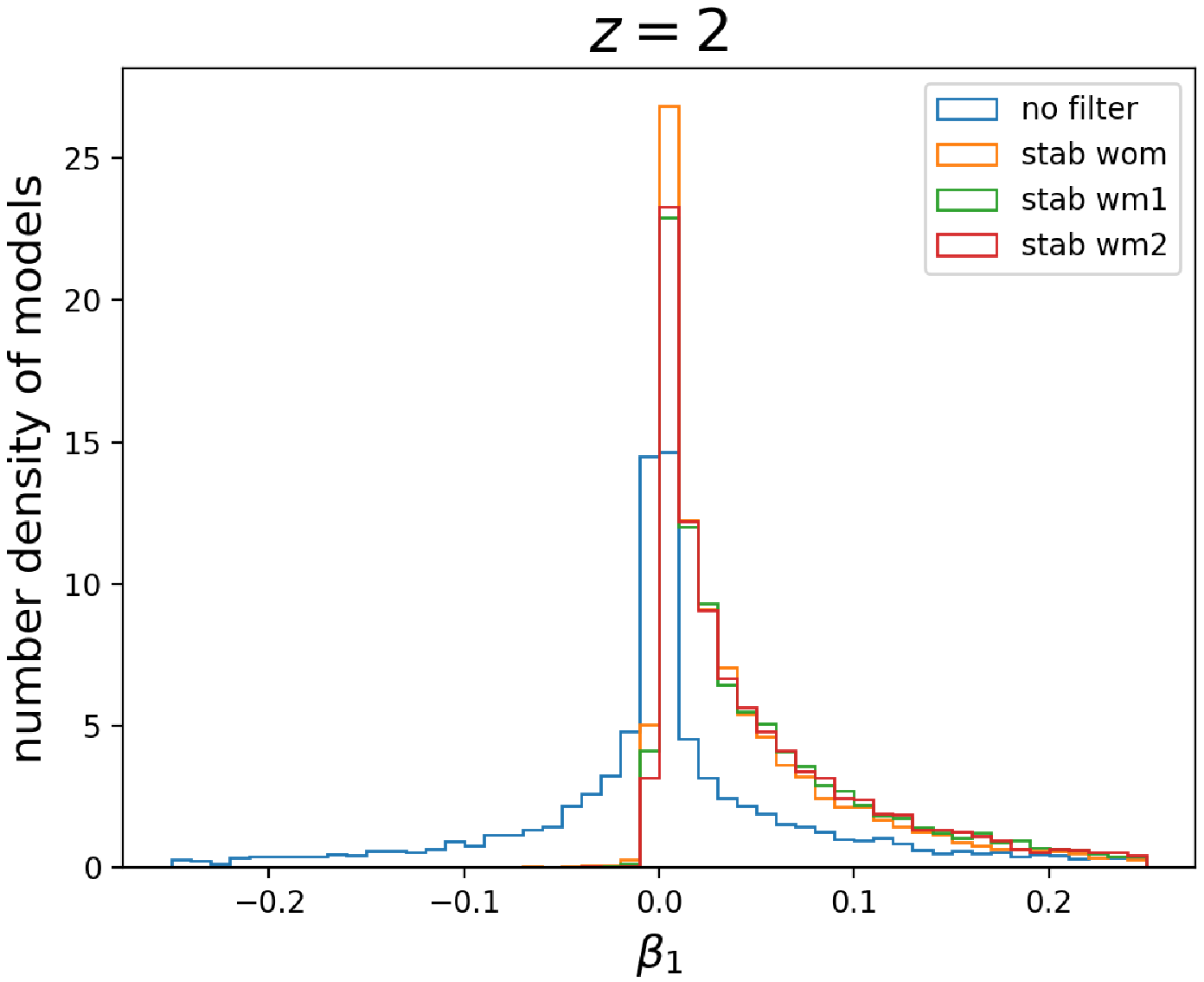}
    \caption{Model distribution in $\beta_1$ for different filters in the redshift $z=0$ (top) and $z=2$ (bottom).} 
   \label{fig:gauge_dep}
\end{center}
\end{figure}

Here we show how a choice of the basis for the cosmological perturbation affects the observables that we are interested in. We pick up three different choices of the bases; {\it stab wom},{\it stab wm1}, and {\it stab wm2}, and their respective stability conditions are

\begin{align}
&{\rm stab\ wom}: A_{\tilde{\zeta}}>0,\ B_{\tilde{\zeta}}<0\,,\ M^2_*>0\,,\label{stab_wom}\\
&{\rm stab\ wm1}: A_{\tilde{\zeta}} >0,\  B_{\tilde{\zeta}}+\frac{\rho_m+p_m}{M^2_*H^2}\left( \frac{1+\alpha_H+\beta_1}{1+\alpha_B-\dot{\beta}_1/H}\right)^2 < 0\,,\ M^2_*>0 \,,\label{stab1}\\
&{\rm stab\ wm2}: A_{\tilde{\zeta}}+\frac{\rho_m+p_m}{M^2_*H^2} \frac{3 \beta_1(2+3c^2_m\beta_1)}{(1+\alpha_B-\dot{\beta}_1/H)^2}>0\,,\ \cr 
& \qquad \qquad \qquad B_{\tilde{\zeta}}+\frac{\rho_m+p_m}{M^2_*H^2}\left( \frac{1+\alpha_H+\beta_1}{1+\alpha_B-\dot{\beta}_1/H}\right)^2 < 0\,,\ M^2_*>0 \,.\label{stab2}
\end{align}
Fig.~\ref{fig:gauge_dep} provide how the three filtering methods for the stability conditions affect the posterior distribution of the characteristic parameters. We used the above stability conditions to filters the models and plotted posterior distributions in Fig.~\ref{fig:gauge_dep}. By comparing the top and bottom figure of Fig. \ref{fig:gauge_dep}, we confirm that the distribution of the characteristic parameters are unaffected by the choice of the basis in our interested redshift range. 

In the deep matter dominant or radiation dominant epoch the basis may severely affect the stability conditions. In fact, the additional terms appearing in the stability coefficients without the matter could be compatible in the matter dominant epoch, namely $\rho_m/3M^2_*H^2$. 

We might need a more sophisticated and careful in stability analysis of those epochs. This point, however, is beyond the scope of this paper. Hence, we conclude that a choice of the basis for the stability condition of the scalar and the matter fluctuation is less important for our late time Universe up to $z=2$ for the assumptions of the slow-rolling scalar field. In this paper, we have used the stability conditions obtained in the ref. \cite{Crisostomi:2019yfo} as our stability filter to obtain the results in section \ref{sec:modeldist}.

%%%%%%%%%% End Appendix %%%%%%%%%%%%%
%%%%%%%%%%%%%%%%%%%%%%%%%%%%%%%%%%%%%

%%%%%%%%%%%%%%%%%%%%%%%%%%%%%%%%%%%%%
%%%%%%%%%% Bibliography %%%%%%%%%%%%%
\bibliography{bibliography} % Produces the bibliography via BibTeX.

%%%%%%%%%%%%%%%%%%%%%%%%%%%%%%%%%%%%%
\end{document}